\documentclass[10pt,journal,compsoc]{IEEEtran}
\ifCLASSOPTIONcompsoc
  \usepackage[nocompress]{cite}
\else
  \usepackage{cite}
\fi
\ifCLASSINFOpdf
\else
\fi

\usepackage{algorithmicx}
\usepackage{algorithm}
\usepackage{array}
\usepackage{textcomp}
\usepackage{stfloats}

\usepackage{algorithmicx}
\usepackage{algorithm}
\usepackage{array}
\usepackage{textcomp}
\usepackage{stfloats}

\usepackage{amsmath,amssymb,amsfonts}

\usepackage{xcolor}
\usepackage{times}
\usepackage{xcolor}
\usepackage{soul}
\usepackage{multirow}
\usepackage{subfigure}
\usepackage[utf8]{inputenc}
\usepackage{diagbox}
\usepackage{bm}

\usepackage{algpseudocode}

\newtheorem{lemma}{\textbf{Lemma}} 
\newtheorem{theorem}{\textbf{Theorem}}
\newtheorem{assumption}{\textbf{Assumption}} 
\newtheorem{definition}{\textbf{Definition}} 
\newtheorem{remark}{\textbf{Remark}}

\usepackage{url}
\usepackage{verbatim}
\usepackage{graphicx}
\usepackage{cite}
\begin{document}
\title{\huge HiFlash: Communication-Efficient Hierarchical Federated Learning with  Adaptive Staleness Control and Heterogeneity-aware Client-Edge Association}

\author{Qiong Wu, Xu Chen,~\IEEEmembership{Senior Member,~IEEE}, Tao Ouyang, Zhi Zhou,~\IEEEmembership{Member,~IEEE}, Xiaoxi Zhang,~\IEEEmembership{Member,~IEEE}, Shusen Yang,~\IEEEmembership{Senior Member,~IEEE}, and Junshan Zhang,~\IEEEmembership{Fellow,~IEEE}
\thanks{Q. Wu, X. Chen, T. Ouyang, Z. Zhou and X. Zhang are with School of Computer Science and Engineering, Sun Yat-sen University, Guangzhou 510006, China. 
	
Shusen Yang is with School of Mathematics and Statistics, Xi'an Jiaotong University, Xi'an, 710049, China. 

Junshan Zhang is with the ECE Department, University of California Davis, United States. 
	}

}

\IEEEtitleabstractindextext{%
\begin{abstract}
Federated learning (FL) is a promising paradigm that enables collaboratively learning a shared model across massive clients while keeping the training data locally. However, for many existing FL systems, clients need to frequently exchange model parameters of large data size with the remote cloud server directly via wide-area networks (WAN), leading to significant communication overhead and long transmission time. To mitigate the communication bottleneck, we resort to the hierarchical federated learning paradigm of HiFL, which reaps the benefits of mobile edge computing and combines synchronous client-edge model aggregation and asynchronous edge-cloud model aggregation together to greatly reduce the traffic volumes of WAN transmissions. Specifically, we first analyze the convergence bound of HiFL theoretically and identify the key controllable factors for model performance improvement. We then advocate an enhanced design of HiFlash by innovatively integrating deep reinforcement learning based adaptive staleness control and heterogeneity-aware client-edge association strategy to boost the system efficiency and mitigate the staleness effect without compromising model accuracy. Extensive experiments corroborate the superior performance of HiFlash in model accuracy, communication reduction, and system efficiency.
\end{abstract}

\begin{IEEEkeywords}
Federated learning, hierarchical mechanism, staleness control, client-edge association.
\end{IEEEkeywords}}

\maketitle

\IEEEdisplaynontitleabstractindextext

% \fi
\IEEEpeerreviewmaketitle

\IEEEraisesectionheading{\section{Introduction}\label{sec:introduction}}
\label{SectionIntroduction}
\IEEEPARstart{N}{owadays}, federated learning (FL) has gained growing attention as it collaboratively trains a global machine learning (ML) model in distributed manner without exposing the data from private clients \cite{mcmahan2017communication, kairouz2021advances}. During the training procedure of FL, the (local/global) model updates are iteratively exchanged between clients and the cloud server until reaching a desirable accurate model, thus it achieves a privacy-preserving learning by leaving training data on local clients. Various popular AI applications such as computer vision \cite{liu2020fedvision}, language processing \cite{yang2018applied} and human activity recognition \cite{chen2019communication} have been derived within this framework.

For many existing FL systems, regardless of synchronous update (e.g., FedAvg \cite{mcmahan2017communication} and its variants \cite{wang2018edge, sahu2018convergence}) or asynchronous update (e.g., FedAsync \cite{Xie2019Asyn}), massive model parameters need to be exchanged in multiple update iterations. However, clients geographically scattered over the edges of networks are usually connected to a remote cloud server through wide-area networks (WAN) and long-distance transmissions, which would incur high communication cost and serious network congestion. Such communication inefficiency would greatly deteriorate the system performance of large-scale distributed training and further hinder the wide deployment of FL systems in practice. Hence, the research issue of boosting the communication efficiency of FL has recently drawn great attention\cite{lim2020federated}.

Hierarchical architecture is a promising solution to alleviate the huge communication pressure of the cloud server, since an order of magnitude fewer data-size of model update would be transferred to cloud by aggregating local models at the lower layer in advance. Due to the merits of mobile edge computing (MEC) in practice, edge nodes (e.g., 5G edge servers) can be set as the intermediates for local model aggregation  \cite{wu2020accelerating}. The rationales are as follows: 1) due to shorter routing path and less hop distance in the local-area network (LAN), a lower network delay and reduced network jitter are offered in the edge layer \cite{XuFMZLQWLYL21}. Further, the straggler problem caused by less effective communication between cloud and clients can be significantly alleviated; 2) compared to high monetary cost of WAN usage in traditional FL, abundant cheaper LAN resources at the edge nodes promote FL deployment in reality \cite{yuan2020hierarchical}; 3) FL applications are commonly scattered over massive devices, which are naturally clustered into many edge domains (e.g., campus and hospital) \cite{9296274}. This distributed pattern can be well accommodated in hierarchical FL. 

Motivated by these facts, a new paradigm of client-edge-cloud hierarchical FL has recently been put forward \cite{liu2020client}, \cite{luo2020hfel}, which involves two levels of synchronous model aggregations, i.e., client model aggregation controlled by the edge nodes at lower layer, edge model aggregation controlled by the cloud server at higher layer. This framework aims at leveraging the advantage of synchronous update to train global model with high accuracy and fast convergence at the lower layer, benefited from high LAN bandwidth and sufficient computation resources at the edge nodes. However, a severe straggler problem at the higher layer would be incurred due to the edge heterogeneity (e.g., diverse WAN connection conditions and heterogeneous edge aggregation time due to different client size) and the communication bottleneck from edge to cloud. More explicitly, large waiting time in synchronous global model aggregation at the higher layer is inevitable.

\begin{figure}[!t]
	\centering
	\includegraphics[width=0.85\linewidth]{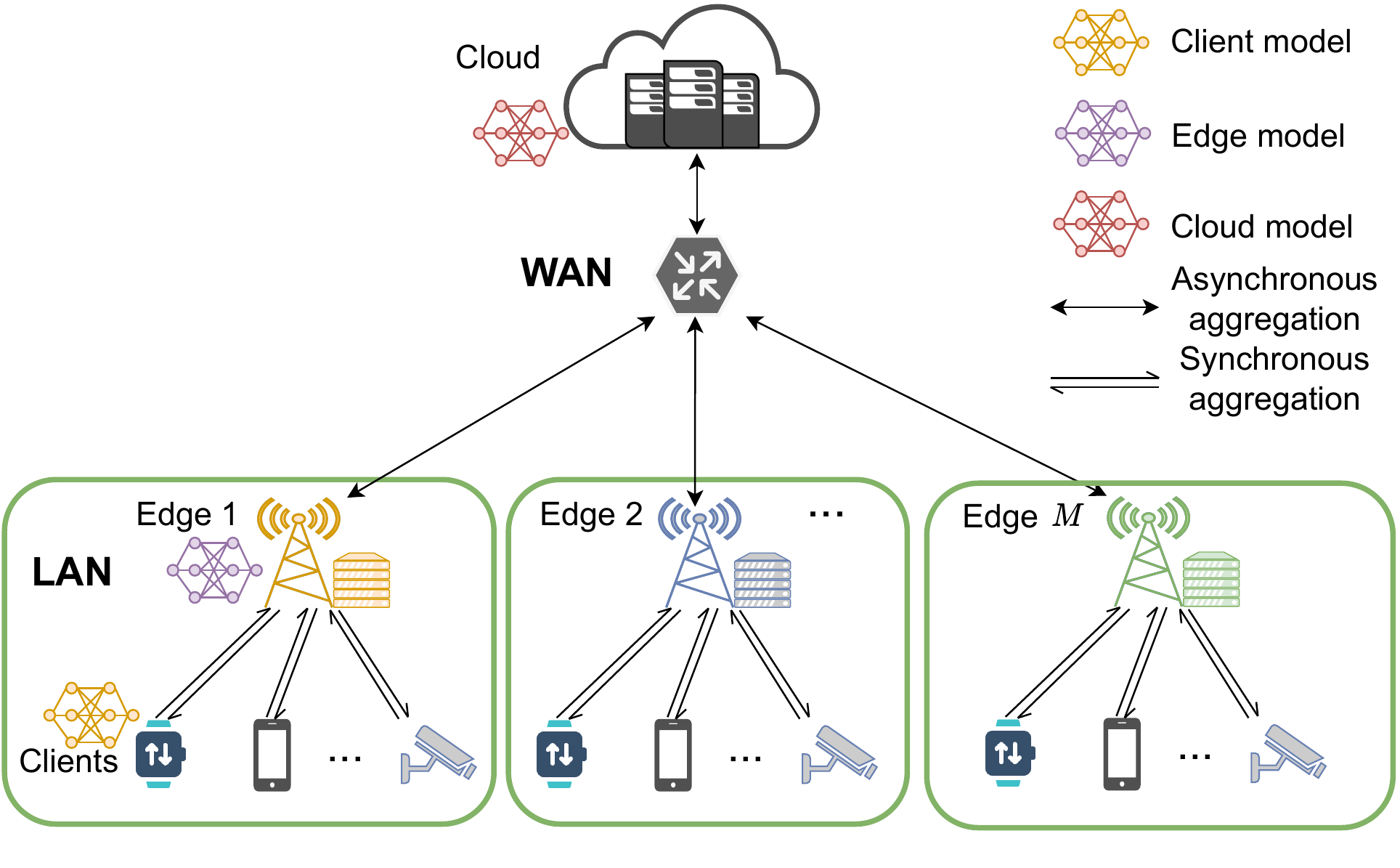}
	\vspace{-10pt}
	\caption{Overview of HiFL approach.}
	\label{HiFL_overview}
	\vspace{-10pt}
\end{figure}

To fully unleash the benefits of hierarchical FL, in this paper we incorporates the merits of synchronous and asynchronous operations in different aggregation layers into the hierarchical FL, which we call HiFL in order to differentiate it from HierFAVG, the version of hierarchical FL with two levels of synchronous model aggregations. As depicted in Fig. \ref{HiFL_overview}, confronted with huge edge heterogeneity and complicated WAN environment among edge nodes and cloud, asynchronous update is adopted for edge-cloud model aggregation to improve learning efficiency via wait-free communication. At the lower layer, synchronous model aggregation between clients and edge nodes ensures high accuracy and fast convergence. Moreover, benefited from high LAN bandwidth and sufficient computation resources at the edges in the communication-efficient one-hop access edge network environment, the straggler problem is very mild and can be neglected during synchronous client-edge aggregation, compared with the asynchronous edge-cloud aggregation communications over the latency-significant WAN.

Nevertheless, HiFL also brings in new challenges on account of the asynchronous aggregation and hierarchical mechanism design. On one hand, staleness effect arised in asynchronous update negatively impacts on the model accuracy and convergence speed \cite{chai2020fedat}. Existing staleness-tolerant mechanisms usually dampen the impacts of stale model updates by only controlling the trade-off between convergence rate and variance reduction according to the staleness \cite{Xie2019Asyn}. However, its impact on system efficiency (e.g., training time, resource efficiency) is much less considered. For example, a model with large staleness may marginally contribute to the global model, which results in more rounds of communication to reach a target accuracy for asynchronous FL. Thus, it is critical to control the model staleness for communication-efficient model learning. On the other hand, the hierarchical mechanism introduces data heterogeneity among edges, which can be further amplified by the hierarchical client-edge-cloud model aggregation and lead to degraded model performance \cite{chai2020fedat}. Besides, the resource heterogeneity among the edge-associated clients can exacerbate the straggler effects, possibly prolonging the waiting time in client-edge model aggregation.

To cope with the above challenges, in this paper we first investigate HiFL to gain useful theoretical insights about its performance bound and system efficiency, and then identify the key controllable parameters that affect the learning performance. Motivated by our theoretical results, we devise an adaptive staleness control strategy for edge-cloud layer and a heterogeneity-aware association mechanism for client-edge layer to improve the overall efficiency of HiFL. For staleness control, existing approaches usually assume a pre-defined fixed threshold for the participating clients, which can not well adapt to the realistic dynamic environment. Moreover, the threshold determination is non-trivial due to the complicated FL environments (e.g., data and resource heterogeneity of clients, current running stages of FL model). Differently, we resort to the deep reinforcement learning (DRL) method and design a DRL agent based on Deep Q-Network (DQN) \cite{mnih2013playing} to wisely make adaptive staleness threshold decisions tailored to the dynamic and complicated FL environments. The DRL agent is trained through a Double DQN for increased efficiency and robustness. For client-edge association, we devise an efficient weighted heuristic to find a near optimal solution that jointly minimizes the data heterogeneity among the edges and resource heterogeneity in the edge-associated clients.

In summary, this paper makes the following contributions:
\begin{itemize}
	\item To achieve communication-efficient and accurate model learning, we resort to HiFL, a hierarchical federated learning approach that performs synchronous client-edge model aggregation and asynchronous edge-cloud model aggregation. Rigorous theoretical analysis for the convergence of HiFL is provided, including both convex and non-convex learning objectives.
	\item Inspired by the theoretical convergence analysis, we further advocate an enhanced design of HiFlash, which introduces adaptive staleness control and heterogeneity-aware client-edge association based on HiFL. The HiFlash approach enables large-scale  deployment with boosted model performance and system efficiency.
	\item We devise a DRL agent based on a Deep Q-Network (DQN) for adaptive staleness control with elaborative learning reward design in order to improve system efficiency without compromising model accuracy. To mitigate the accuracy degradation and straggler effect caused by data and resource heterogeneity, we establish an efficient weighted heuristic of low-complexity for client-edge association that well balances the trade-off between model accuracy and system efficiency.
	\item Extensive experiments are conducted using three widely adopted image classification datasets to evaluate the effectiveness of HiFlash, demonstrating that HiFlash significantly outperforms other FL based approaches in communication efficiency without compromising model accuracy. For example, even under highly skewed data distributions among clients, HiFlash can still achieve a high model accuracy, and meanwhile greatly reduces communication overhead, e.g., with a reduction ratio of $42\%$ and $89\%$ over the benchmarks of HierFAVG and FedAvg, respectively. 
\end{itemize}
The rest of this paper is organized as follows: Section \ref{SectionPreliminaries} presents the preliminaries on FL and DRL. Section \ref{SectionHiFL} introduces a hierarchical FL approach named HiFL. In Section \ref{SectionConvergenceAnalysis}, we provide theoretical analysis for HiFL, and further devise HiFlash, an enhanced HiFL with adaptive staleness control and heterogeneity-aware client-edge association in Section \ref{SectionHiFlash}. Extensive experiments are conducted in Section \ref{SectionExperiments}. We review the related work in Section \ref{SectionRelatedWork} and conclude the paper in Section \ref{SectionConclusion}.

\section{Preliminaries}
\label{SectionPreliminaries}
\subsection{Federated Learning}
Federated learning \cite{mcmahan2017communication}, first proposed by Google in 2016, trains a global shared model among massive clients in a privacy-preserving manner, where a central server serving as an aggregator coordinates client model learning. In general, a FL system consists one cloud server and $N$ dispersed clients. Each client $k$ has a collection of local dataset $\mathcal{D}_{k}=\{\mathbf{x}_j,y_j\}_{j=1}^{|\mathcal{D}_k|}$, where $\mathbf{x}_j$ is the feature of training sample $j$ and $y_j$ is its ground-truth label. To collaboratively train a global ML model $\omega \in \mathbb{R}^{d}$, its loss function associated with the data sample $(\mathbf{x}_j,y_j)$ is denoted as $f(\mathbf{x}_j,y_j,\omega)$, where $d$ is the total number of trainable parameters. For ease of exposition, we use $f_j(\omega)$ to replace $f(\mathbf{x}_j,y_j,\omega)$ notation. As a result, the learning objective of FL is to minimize the loss function over the collection of training data at $N$ clients, i.e.,
\begin{equation}
	\small
	\label{2FLobjective}
	\min_{\omega \in \mathbb{R}^{d}} F(\omega) = \sum_{k=1}^{N}\frac{|\mathcal{D}_k|}{|\mathcal{D}|}F_k(\omega), 
\end{equation}
where $F_{k}(\omega) = \frac{1}{|\mathcal{D}_{k}|}\sum_{j\in \mathcal{D}_{k}}f_{j}(\omega)$ is the loss empirical objective over the data samples at client $k$, which is task-specified, for example, the learning objective can be cross-entropy loss for image classification tasks. Assuming $\mathcal{D}_k \cap \mathcal{D}_{k^{\prime}} = \emptyset$ for $k \neq k^{\prime}$, we define $\mathcal{D} = \cup_{k=1}^N\mathcal{D}_k$ and use $|\cdot|$ to denote the size of a set. 

To solve the optimization problem in Eqn. (\ref{2FLobjective}), FedAvg, the most widely used FL framework, proposes to run local stochastic gradient descent (SGD) in parallel on a sampled subset of clients and conducts synchronous model aggregation via a central server once in a while \cite{mcmahan2017communication}. The process is repeated until the model reaches a desired accuracy. Due to slow and expensive network connection (e.g., frequent backhaul and long communication distance) between the cloud server and the geographically distributed clients \cite{wu2020accelerating}, FedAvg performs multiple local learning steps before uploading the model updates into the cloud server, so that the number of communication rounds is considerably reduced and the network burden is further relieved.

\subsection{Deep Reinforcement Learning}
In reinforcement learning (RL), a RL agent interacts with the environment in discrete time slots to maximize its reward in the long run. At each time slot $i$, the RL agent observes state $\mathbf{s}^{i}$, executes action $a^{i}$ and receives a reward $r^{i}$ from the environment. The state $\mathbf{s}^{i}$ of the environment then transits to $\mathbf{s}^{i+1}$ for the action decision making of next time slot. The whole process follows a Markov Decision Process (MDP) \cite{sutton2018reinforcement} which can be described as a tuple $\mathcal{M} = <\mathcal{S},\mathcal{A},\mathcal{P},\mathcal{R},\gamma>$ wherein $\mathcal{S}$ is the state space, $\mathcal{A}$ is the action space and $\mathcal{P}: \mathcal{S} \times \mathcal{A} \times \mathcal{S} \rightarrow [0,1]$ is a probabilistic transition function. $\mathcal{R}$ is the immediate reward function and $\gamma \in [0,1]$ is a factor discounting the future rewards. The objective of the RL agent is to learn a policy $\pi^{*}$, a mapping between states and actions that maximizes the cumulative discounted reward $R = \sum_{i=1}^{I}\gamma^{i-1}r^{i}$, where $I$ is the total running slots.

To estimate the expected cumulative discounted reward starting from state $\mathbf{s}^{i}$, value-based RL approaches adopt an action-value function
\begin{equation}
	\small
	\label{action_value_function}
	\begin{aligned}
		Q_{\pi}(\mathbf{s}^{i},a^{i}) &= \mathbb{E}_{\pi}[\sum_{\hat{i}=1}^{\infty}\gamma^{\hat{i}-1}r^{i+\hat{i}-1}|\mathbf{s}^{i},a^{i}]\\
		&=\mathbb{E}_{\pi}[r^{i} + \gamma Q_{\pi}(\mathbf{s}^{i+1},a^{i+1})|\mathbf{s}^{i},a^{i}],
	\end{aligned}
\end{equation}
where $\pi$ is the state-action mapping policy. The optimal action-value function $Q^{*}(\mathbf{s}^{i},a^i)$ is defined as the maximum expectation of the cumulative discounted reward:
\begin{equation}
	\small
	\label{action_value_function2}
	Q^{*}(\mathbf{s}^{i},a^i) =\mathbb{E}_{\pi}[r_i + \gamma \max_{a} Q^{*}(\mathbf{s}^{i+1},a)|\mathbf{s}^{i},a^i].
\end{equation}
Hence, we could apply function approximation techniques to learn the action-value funtion $Q_{\pi}(\mathbf{s}^{i},a^{i},\theta_{i})$ approximating the optimal function $Q^{*}$.

Nevertheless, for many real-world problems, the state space becomes too large to keep track of all the Q-values. To alleviate this issue, deep reinforcement learning (DRL) proposes to adopt DNN as the approximator of the action-value function $Q_{\pi}$ by leveraging the powerful generalization abilities of DNNs. For example, Deep Q-Network (DQN) \cite{mnih2013playing} uses a DNN to estimate the Q-values of states and actions, and the objective of DQN is minimizing the mean-squared error (MSE) loss between the target $r^{i} + \gamma \max_{a^{i+1}} Q_{\pi}(\mathbf{s}^{i+1},a^{i+1},\theta_{i})$ and the approximator described as follows:
\begin{equation}
	\small
	\label{DQN_objective}
	\arg \min_{\theta_{i}} \mathbb{L}(\theta_{i})=(r^{i} + \gamma \max_{a^{i+1}} Q_{\pi}(\mathbf{s}^{i+1},a^{i+1},\theta_{i}) - Q_{\pi}(\mathbf{s}^{i},a^{i},\theta_{i}))^2.
\end{equation}
For ease of convergence, DQN transforms DRL as a form of supervised learning and induces experience replay \cite{mnih2015human}, which contains abundant transition samples, for correlation reduction between samples.

\begin{figure}[!t]
	\centering
	\includegraphics[width=0.86\linewidth]{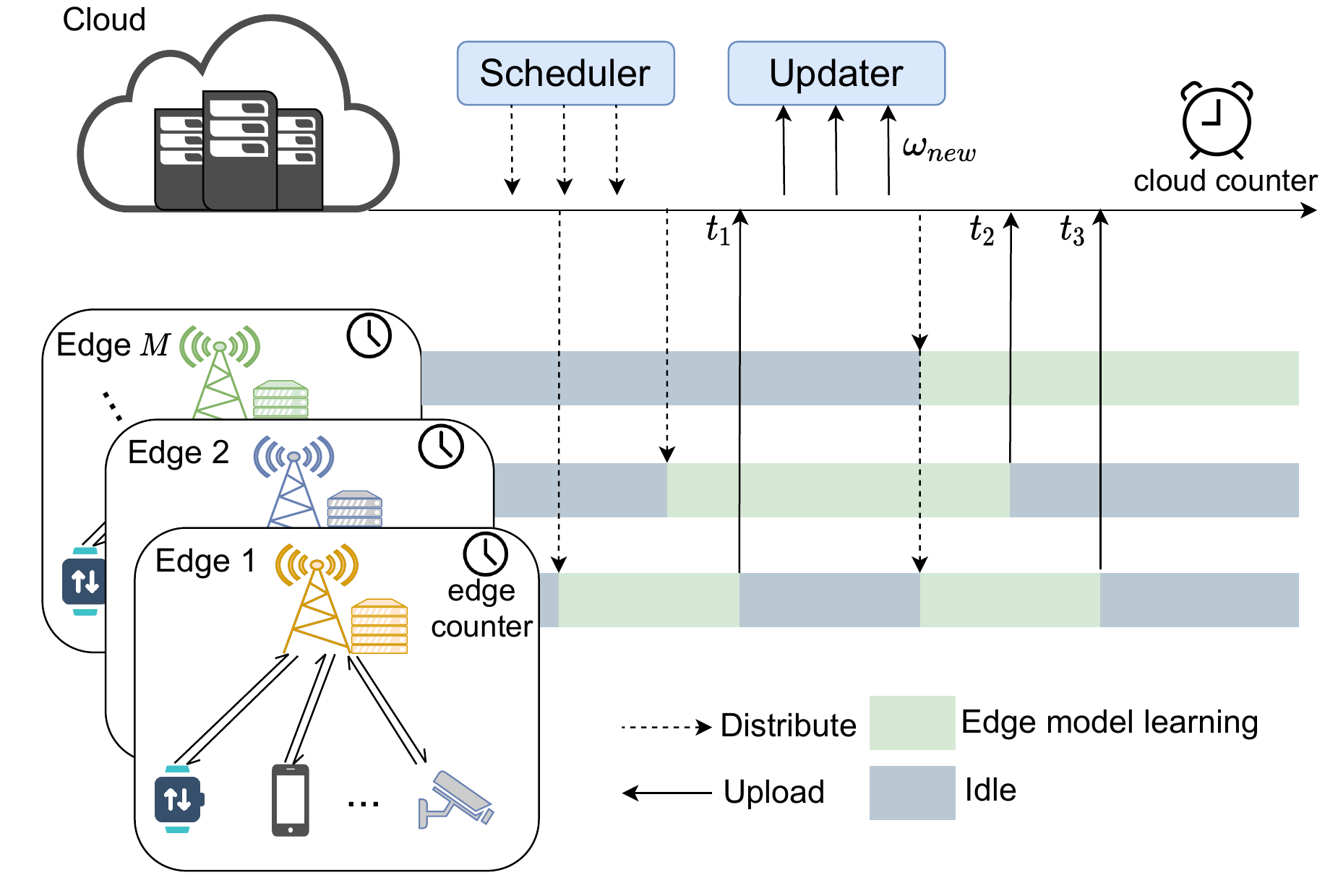}
	\vspace{-10pt}
	\caption{Model training procedure of HiFL.}
	\label{HiFL}
	\vspace{-10pt}
\end{figure}

\section{HiFL Design}
\label{SectionHiFL}
\subsection{Problem Definition and HiFL Overview}
We first provide the problem definition of HiFL based on the traditional FL. As depicted in Fig. \ref{HiFL}, client model updates are not directly sent to the cloud but edge nodes. More explicitly, $N$ participant clients are divided into $M$ disjoint groups  based on their characteristics (e.g., geographical locations), each of which is associated with one edge node. In general, the number of edge nodes is far less than clients, i.e., $M \ll N$. We denote $\mathcal{C}^{m}$ as the client set of edge node $m$, and total participant clients can be defined as $\mathcal{N} = \cup_{m=1}^M\mathcal{C}^m$. Thus, based on this hierarchical FL architecture, the learning objective in Eqn. (\ref{2FLobjective}) is extended as:
\begin{equation}
	\small
	\label{3FLobjective}
	\min_{\omega \in \mathbb{R}^{d}} F(\omega) = \sum_{m=1}^{M}\frac{|\mathcal{D}^m|}{|\mathcal{D}|}F^m(\omega), 
\end{equation}
where $F^m(\omega) = \sum_{k\in \mathcal{C}^m}\frac{|\mathcal{D}_{k}|}{|\mathcal{D}^m|}F_{k}(\omega)$ denotes the objective on edge node $m$, which is a linear combination of the empirical objectives of clients in $\mathcal{C}^m$. $|\mathcal{D}^{m}| =  \sum_{k \in \mathcal{C}^{m}}|\mathcal{D}_{k}|$ is the data size of all samples across the clients associated with edge node $m$. Table \ref{notation} lists the key notations in our paper.

Based on the disparate behaviors of cloud and edges, HiFL adopts synchronous client-edge model aggregation and asynchronous edge-cloud model aggregation to synergistically train a high-quality ML model in a cost-efficient way. At the side of edge, high LAN bandwidth and reduced network jitter considerably shorten the communication latency for gradient exchanges or model download. Within the same LAN environment, synchronous model aggregation is more desired between edge nodes and clients, due to its high model training precision and fast convergence speed. However, at the side of cloud, model training suffers from communication bottleneck in the complicated WAN environment (e.g., highly fluctuating long-distance transmission time and diverse edge model aggregation time due to different size of clients with different edges), which leads to severe straggler problem. Thus, asynchronous aggregation is adopted to mitigate this straggler effect via reducing the waiting time of model updates between the central server and the edge nodes.

\begin{table}[!t]
	%\normalsize	
	\caption{List of key notations. }
	\vspace{-5pt}
	\newcommand{\tabincell}[2]{\begin{tabular}{@{}#1@{}}#2\end{tabular}}
	\label{notations}       % Give a unique label
	\centering
	\begin{tabular}{|c|c|}
		\hline
		Symbol &  Description\\
		\hline \multicolumn{2}{|l|}{\textit{\textbf{In General Federated Learning Settings}}}{}\\
		\hline $N$ & the number of clients\\
		\hline $M$ & the number of edges\\
		\hline $\mathcal{D}_k$ & the dataset of client $k$\\
		\hline $\mathcal{C}^m$ & the set of clients associated with edge $m$\\
		\hline $t_{e}$, $t_{c}$ & edge counter and cloud counter\\
		\hline $\eta$& learning rate of federated learning\\
		\hline $\omega(t_{c})$ & the cloud model at $t_{c}$-th global iteration\\
		\hline $\omega^m(t_{c},t_{e})$ & edge model at $t_{e}$-th iteration computed based on $\omega(t_{c})$\\
		\hline $\omega_k(t_c,t_e)$& client model at $t_{e}$-th iteration computed based on $\omega(t_{c})$\\
		\hline \multicolumn{2}{|l|}{\textit{\textbf{In Client-edge Model Aggregation}}}{}\\
		\hline $H$ & the number of client-edge model aggregation\\ 
		\hline $c$ & \tabincell{c}{the number of learning epochs on the client \\before synchronizing with the edge node}\\
		\hline \multicolumn{2}{|l|}{\textit{\textbf{In Edge-cloud Model Aggregation}}}{}\\
		\hline $\tau$ & the staleness of the edge model\\
		\hline $\alpha$, $\upsilon$ & \tabincell{c}{initial model weight of the edge model and penalty\\ coefficient for calculating mixing hyperparameter \\$\alpha_{\tau}$ in cloud model aggregation}\\			
		\hline \multicolumn{2}{|l|}{\textit{\textbf{In Adaptive Staleness Control}}}{}\\
		\hline $\mathbf{s}^{i}$, $a^{i}$ & the state and action of DRL agent in time slot $i$\\
		\hline $r^i$ & total system cost at time slot $i$\\
		\hline $\sigma_{1}$, $\sigma_{2}$ & penalty factors for computation and communication costs\\
		\hline \multicolumn{2}{|l|}{\textit{\textbf{In Heterogeneity-aware Association}}}{}\\
		\hline $\lambda$ & weighting parameter for data and resource heterogeneity\\
		\hline $L^{m,k}$ & response latency for client $k$ associated with edge $m$\\
		\hline
	\end{tabular}
	\vspace{-10pt}
	\label{notation}
\end{table}

As shown in Fig. \ref{HiFL}, we design two core components, i.e., scheduler and updater, running asynchronously in parallel on the cloud server to achieve the wait-free goal, where the former one is in charge of latest model distribution (which can be integrated with control functionality by DRL agent specified later on), and the latter one is for global model aggregation. More explicitly, once an idle edge node gets engaged in model training for its interest, it will actively inform the cloud server to download the latest version of global model. Then the cloud server will check its updater and immediately send the result to corresponding edge node. Receiving the global model, the edge node quickly broadcasts it to the associated clients and leverages their local datasets to collaboratively train a shared edge model in a synchronous manner with efficient client-edge communication. If the cloud server receives a trained model from an edge node, the updater will update the global model immediately, without waiting for other edge nodes. In order to control the model staleness caused by asynchronous aggregation, the updater conducts model aggregation with a weight penalty on the received model update, which will be elaborated in Section \ref{TraingProcess}. 

Note that to improve the throughput of HiFL, multiple edge model training processes can be executed in parallel, which results in multiple updater threads with read-write lock on the global model. 
This asynchronous model aggregation strategy relieves the network congestion on the cloud side and enables wait-free global model learning, further reducing the communication overhead and speeding up the model training process. Particularly, different counters are set to record the model update times in this hierarchical settings, since the clients update the edge model without the coordination of the cloud server. Thus, we design cloud counter $t_{c}$ and edge counter $t_e$ for asynchronous cloud model aggregation and synchronous edge model aggregation, respectively.

\subsection{HiFL Training Process}
\label{TraingProcess}
The learning process of HiFL contains two main procedures, including 1) client-edge model aggregation, and 2) edge-cloud model aggregation, as elaborated below.

\textbf{1) Synchronous client-edge model aggregation.} When edge node $m$ receives current global model $\omega(t_c)$ from the cloud server, the edge model is initialized as $\omega^m(t_c,t_e) = \omega(t_c)$ with $t_e = 0$, where $t_c$ indicates the number of global aggregations conducted on the cloud model, $t_e$ denotes the number of local model updates for edge model aggregation after receiving cloud model $\omega(t_c)$ \footnote{The maximum value of $t_c$ can be determined by the cloud server based on the convergence of the model, while the maximum value of $t_e$ can be set by each edge node and hence is different across various nodes.}. The edge model is further sent to the associated clients for client model update. 

We adopt the widely used FedAvg algorithm \cite{mcmahan2017communication} to collaboratively train a satisfactory edge model. Specifically, at $t_e$-th iteration, client $k \in \mathcal{C}^m$ performs model update with its local data. To reduce communication overhead, the classical FedAvg method aggregates all client models associated with edge node $m$ and synchronizes with edge model $\omega^m$ after every $c$ steps of local updates on each client. Denote $\omega_{k}^{m}(t_c,t_e)$ as the model parameters of client $k \in \mathcal{C}^m$, then $\omega_{k}^{m}(t_c,t_e)$ evolves in the following way:
\begin{small}
	\begin{equation}
		\label{clientModel}
		\omega_{k}^{m}(t_c,t_e) \!=\! \begin{cases}
			\omega_{k}^{m}(t_c,t_e\!-\!1)\!-\!\eta\nabla F_k(\omega_{k}^{m}(t_c,t_e\!-\!1)), &\!\! t_e\!\!\!\!\!\mod\! c\!\neq\! 0\\
			\omega^{m}(t_c, t_e),&\!\! t_e\!\!\!\!\!\mod\! c \!=\! 0
		\end{cases}
	\end{equation}
\end{small}
where 
\begin{small}
	\begin{equation}
		\label{clusterModel}
		w^m(t_c,t_e) = \sum_{k \in \mathcal{C}^{m}} \frac{|\mathcal{D}_{k}|[\omega_{k}^{m}(t_c,t_e\!-\!1)-\eta\nabla F_k(\omega_{k}^{m}(t_c,t_e\!-\!1))]}{|\mathcal{D}^{m}|}.
	\end{equation}
\end{small}
Without loss of generality, we assume that the edge node performs a number of $H$ model aggregations (e.g., $t_e \le Hc$) in each global training round, which indicates that $Hc$ client model updates have been performed for one client during the client-edge model aggregation. After this collaborative model training, the edge model is updated as $w^m(t_c,Hc)$ and will asynchronously update the global model with the cloud server.

\textbf{2) Asynchronous edge-cloud model aggregation.} The asynchronous mechanism in edge-cloud model aggregation introduces the challenge of staleness as multiple edge nodes are free to perform model training and uploading at arbitrary times. For example, at the global counter $t_c$, the cloud server receives a stale model $\omega^m(t_c-\tau,Hc)$ which is trained by edge node $m$ based on the global model $\omega(t_c-\tau)$, where $\tau$ represents the staleness of the edge model. As the edge model is trained based on an outdated cloud model version, the stale model will add noise to the cloud model training procedure, slow down or even prevent the training convergence \cite{damaskinos2020fleet}.

To control the error caused by asynchrony, HiFL updates the global model with the stale edge model by introducing a mixing hyperparameter $\alpha_{\tau}$ as in \cite{Xie2019Asyn},
\begin{equation}
	\label{CloudModelUpdate}
	\omega(t_c) = (1-\alpha_{\tau})\omega(t_c-1) + \alpha_{\tau} \omega^{m}(t_c-\tau,Hc),
\end{equation}
where $\alpha_{\tau}$ is the weight that the edge model $\omega^{m}(t_c-\tau,Hc)$ with staleness $\tau$ contributes to the global model. A smaller $\alpha_{\tau}$ will result in more FL training rounds while a bigger value of $\alpha_{\tau}$ can cause large accuracy fluctuation. By adjusting the value of $\alpha_{\tau}$, we can adaptively control the trade-off between convergence speed and variance reduction in the model learning process. In this paper, we use the following exponential function to determine the value of $\alpha_{\tau}$,
\begin{equation}
	\label{alpha}
	\alpha_{\tau} = \alpha \cdot \upsilon^{\tau},
\end{equation}
where $\alpha \in (0,1)$ is the initial model weight of the edge model. We can decrease $\alpha$ to mitigate the error caused by large staleness $\tau$ with the penalty coefficient $\upsilon \in (0,1)$. 

The cloud server and edge nodes in HiFL conducts model updates asynchronously until the cloud model converges. The synchronous client-edge aggregations on different client groups can be conducted in parallel, and the asynchronous edge-cloud model aggregation avoids from long waiting time, both of which contribute to the fast model learning and wait-free communication. The details of the HiFL algorithm is elaborated in Algorithm \ref{HiFL_algorithm}.

\begin{algorithm}[htb]
	\small
	\caption{HiFL Training Procedure}
	\label{HiFL_algorithm}
	\begin{algorithmic}[1]
		\Require Datasets from $N$ distributed clients $\{\mathcal{D}_{1}, \mathcal{D}_{2},..., \mathcal{D}_{N}\}$, number of local updates $c$, and learning rate $\eta$.
		\State Conduct client clustering based on some predefined criteria (e.g., geographical locations) or by heterogeneity-aware client-edge association strategy in Algorithm \ref{CEA}
		\State \textbf{Cloud server executes}:
		\State \quad Initialize the cloud model $\omega$, $t_c \leftarrow 0$
		\State \quad \textbf{Scheduler:}
		\State \quad \quad Periodically distribute global model $\omega(t_c)$ and global counter $t_c$ to one edge node for edge model update
		\State \quad \textbf{Updater:}
		\State \quad \quad \textbf{for} $t_c = 1, 2, ..., T_c$ \textbf{do}
		\State \quad \quad \quad Receive a pair $(\omega^m(\tilde{t}_{c},Hc), \tilde{t}_{c})$ from one edge node
		\State \quad \quad \quad Calculate the model staleness $\tau = t_c - \tilde{t}_{c}$
		\State \quad \quad \quad Update cloud model with Eqn. (\ref{CloudModelUpdate})
		\State \quad \quad \textbf{end for}
		\State \textbf{Edge node executes}:
		\State \quad \textbf{for} $m \in {1, 2, ..., M}$ in parallel \textbf{do}
		\State \quad \quad \textbf{if} received a pair of the cloud model and the cloud counter $(\omega(t_c),t_c)$ from Scheduler \textbf{then}
		\State \quad \quad \quad Set $\tilde{t}_{c} = t_c$ and $\omega^{m}(\tilde{t}_{c},0) = \omega(t_c)$
		\State \quad \quad \quad Initialize $\omega_k^m(\tilde{t}_{c},0) = \omega^{m}(\tilde{t}_{c},0)$ for $k \in \mathcal{C}^m$
		\State \quad \quad \quad \textbf{for} $t_e = 1,..., Hc$ \textbf{do}
		\State \quad \quad \quad/*Client executes*/
		\State \quad \quad \quad \quad \textbf{for} $k \in \mathcal{C}^{m}$ \textbf{do}
		\State \quad \quad \quad \quad \quad Calculate current client model by Eqn. (\ref{clientModel})
		\State \quad \quad \quad \quad \textbf{end for}
		\State \quad \quad \quad \textbf{end for}
		\State \quad \quad \quad Send $(\omega^m(\tilde{t}_{c},Hc)$,$\tilde{t}_{c})$ to the cloud server.
		\State \quad \quad \textbf{end if}
		\State \quad \textbf{end for}
		\Ensure Cloud model $\omega(T_c)$.
	\end{algorithmic}
	%\vspace{-5pt}
\end{algorithm}

\section{Convergence Analysis}
\label{SectionConvergenceAnalysis}
\subsection{Definitions and Assumptions}
For the purpose of the analysis, we introduce the following definitions and assumptions to the loss function.
\begin{assumption}
	\label{smoothness}
	\textit{(Smoothness).} The function $F_{k}(\omega)$ is $\beta$-smooth if $\forall \omega, \omega^{\prime}$,	
	\begin{equation}
		\small
		\label{smooth2}
		||\nabla F_{k}(\omega) - \nabla F_{k}(\omega^{\prime})|| \le \beta||\omega - \omega^{\prime}||,
	\end{equation}		
	where $\beta > 0$.
\end{assumption}

\begin{assumption}
	\label{strongConvex}
	\textit{(Strong convexity).} The function $F_{k}(\omega)$ is $\mu$-strongly convex if $\forall \omega, \omega^{\prime}$,
	\begin{equation}
		\small
		\label{convex}
		\langle\nabla F_{k}(\omega^{\prime}), \omega - \omega^{\prime}\rangle + \frac{\mu}{2}||\omega - \omega^{\prime}||^{2} \le F_{k}(\omega) - F_{k}(\omega^{\prime}),
	\end{equation}
	where $\mu \ge 0$. Note that if $\mu = 0$, $F_{k}(\omega)$ is convex.
\end{assumption}

\begin{assumption}
	\label{weakConvex}
	\textit{(Weak convexity).} The function $F_{k}(\omega)$ is $\mu$-weakly convex if the function $G_{k}(\omega) = F_{k}(\omega) + \frac{\mu}{2}||\omega||^2$ is convex, where $\mu \ge 0$. Specifically, $F_{k}(\omega)$ is convex if $\mu = 0$ and potentially non-convex if $\mu > 0$.
\end{assumption}

\begin{assumption}
	\label{Lipschitz}
	\textit{(Lipschitz).} The function $F_k(\omega)$ is $\rho$-Lipschitz if $\forall \omega, \omega^{\prime}$,
	\begin{equation}
		\small
		\label{lipschitz}
		||F_{k}(\omega) - F_{k}(\omega^{\prime})|| \le \rho||\omega-\omega^{\prime}||.
	\end{equation}
\end{assumption}

Under these assumptions, Lemma \ref{GroupFuctionProperty} holds for the loss functions of the edge models and the cloud model.

\begin{lemma}
	\label{GroupFuctionProperty}
	$F^{m}(\omega)$ and $F(\omega)$ are $\mu$-strongly convex, $\beta$-smooth and $\rho$-Lipschitz. 
\end{lemma}
\begin{IEEEproof}
	It is straightforward from the aforementioned assumptions, the definition of $F^{m}(\omega)$, $F(\omega)$ and triangle inequality.
\end{IEEEproof}

In the cloud model training process of HiFL, there are two levels of model aggregation, client-edge model aggregation and edge-cloud model aggregation, conducted in parallel. Following \cite{wang2018edge}, we introduce the notion of \textit{virtual cluster model learning} in Definition \ref{definition_virtual} to find the loss divergence between the edge model trained by synchronous client-edge model aggregation and a virtual cluster model where the training data is assumed to exist on a virtual central repository. Next, we formalize \textit{cluster-based gradient divergence} in Assumption \ref{definition_divergence} to characterize the impact of the difference in data distributions across clients and edge nodes on HiFL.

\begin{definition}
	\label{definition_virtual}
	\textit{(Virtual cluster model learning).} For client-edge model aggregation, we use the shorthand notation $[h] = [(h-1)c, hc)$ to indicate an interval between two successive edge model aggregation. Given a certain client cluster $\mathcal{C}^m$ associated with edge node $m$ and the initialized edge model $\omega^m(t_{c},0)=\omega(t_{c})$, for any interval $[h]$, $h = 1, 2, ..., H$, the virtual cluster model $v^{m}_{[h]}(t_{c},t_{e})$ are updated by performing gradient descent on the centralized data examples $\mathcal{D}^m$ owned by $\mathcal{C}^m$, and synchronizes with the federated edge model $\omega^m$ at the beginning of each interval, as shown in Eqn. (\ref{virtualModel}),
	\begin{equation}
		\small
		\label{virtualModel}
		\!\!\!\! v^{m}_{[h]}(t_{c},t_{e}) \!\!= \!\!\begin{cases}
			v_{[h]}^{m}(t_{c},t_{e}\!\!-\!1)\! -\! \eta\nabla F^m(v_{[h]}^{m}(t_{c},t_{e}\!\!-\!1)),\\
			\quad \quad \quad \quad \quad \quad \quad \quad \quad \quad \quad \quad \quad \quad \quad \quad t_{e}\!\!\!\!\!\mod c\neq 0\\
			\omega^m(t_{c},t_{e}), \quad \quad \quad \quad \quad \quad \quad \quad \quad  \quad\quad \; t_{e}\!\!\!\!\!\mod c = 0.
		\end{cases}
	\end{equation}
\end{definition}

\begin{assumption}
	\label{definition_divergence}
	\textit{(Cluster-Based Gradient Divergence).} For any client $k \in \mathcal{C}^{m}$, $\delta_{k}^m$ is assumed as an upper bound of the gradient difference between the local loss function of client $k$ and the edge loss function of edge node $m$, which can be expresses as follows,
	\begin{equation}
		\small
		\label{delta_divergence}
		||\nabla F_k(\omega) - \nabla F^m(\omega) || \le \delta_{k}^{m},
	\end{equation}

Then, we have $\delta^m \triangleq \frac{\sum_{k \in \mathcal{C}^{m}}|D_k| \delta_{k}^{m}}{\sum_{k \in \mathcal{C}^{m}}|D_k|}$ for the client cluster associated with edge node $m$ and $\delta_{max}$ as the biggest gradient difference across $\{\delta^m\}_{m=1}^{M}$.
	
We assume $\Delta$ as an upper bound of the gradient difference between the loss function of any edge node $m$ and that of the global loss function, i.e.,
\begin{equation}
		\small
		\label{Delta_divergence}
		||\nabla F^m(\omega) - \nabla F(\omega) ||^2 \le \Delta.
\end{equation}
We call $\delta_{max}$ as the client-edge divergence and $\Delta$ as the edge-cloud divergence. In addition, the expected squared norm of stochastic gradients on any client $k$ is defined to be uniformly bounded, i.e., 
\begin{equation}
		\small
		\label{V_divergence}
		\mathbb{E}||\nabla F_k(\omega)||^{2} \le V.
\end{equation}
	
For $\mu$-weakly convex loss function $F_{k}$ (which can be non-convex if $\mu >0$), we define $G_{\tilde{\omega}} = F(\omega) + \frac{\tilde{\mu}}{2}||\omega-\tilde{\omega}||^2$ with $\tilde{\mu} > \mu$. Similarly with the convex settings, we assume $G_{\tilde{\omega}}$ is $\tilde{\beta}$-smooth and $\tilde{\rho}$-Lipschitz. $\forall \tilde{\omega} \in \mathbb{R}_d$, we have $||\nabla G_{k,\tilde{\omega}}(\omega) - \nabla G^m_{\tilde{\omega}}(\omega) || \le \tilde{\delta}_{k}^{m}$, $\tilde{\delta}^m \triangleq \frac{\sum_{k \in \mathcal{C}^{m}}|D_k| \tilde{\delta}_{k}^{m}}{\sum_{k \in \mathcal{C}^{m}}|D_k|}$ for the client cluster associated with edge node $m$ and $\tilde{\delta}_{max}$ as the biggest gradient difference across $\{\tilde{\delta}^m\}_{m=1}^{M}$. Furthermore, we assume $||\nabla G^m_{\tilde{\omega}}(\omega) - \nabla G_{\tilde{\omega}}(\omega)||^2 \le \tilde{\Delta}$ and $||\nabla G_{k,\tilde{\omega}}(\omega)||^2 \le \tilde{V}$.
\end{assumption}

\subsection{Convergence of HiFL}
\label{CoA}
Based on the assumptions and definitions above, we have the following convergence guarantees.
\begin{lemma}
	\label{edge_model_divergence_theorem}
	During client-edge model aggregation, for any interval $[h]$ and $t_{e} \in [h]$, we have 
	\begin{equation}
		\small
		\label{diff}
		||\omega^m(t_{c},t_{e}) - v_{[h]}^{m}(t_{c},t_{e})|| \leq g(t_{e}-(h-1)c),
	\end{equation}	
	where 
	\begin{equation}
		\small
		\label{g_def}
		g(x) =  \frac{\delta_{max}}{\beta}((\eta\beta +1)^{x}-1)- \eta\delta_{max}x,  
	\end{equation}
	for any $x = 0,1,2,\cdots$.
\end{lemma}
Furthermore, as $F^m(\cdot)$ is $\rho$-Lipschitz, we have $F^m(\omega^m(t_{c},t_{e})) - F^m(v_{[h]}^{m}(t_{c},t_{e})) \leq \rho g(t_{e}\!-\!(h-1)c)$.

\begin{IEEEproof}
	Please refer to Appendix A of the separate supplementary file for details.	
\end{IEEEproof}
Thus, when the client-edge model aggregation finishes, e.g., $t_{e} = Hc$, the loss divergence between the edge model trained by FL and the virtual cluster model is $F^m(\omega^m(t_{c},Hc)) - F^m(v_{[h]}^{m}(t_{c},Hc)) \leq \rho g(c)$. With the help of the weight deviation upper bound, we are now ready to prove the convergence of HiFL for both convex and non-convex loss functions.

\begin{theorem}
	\label{theorem_convex}
	Suppose the loss function $F_k$ is $\mu$-strongly convex, and each edge node executes $H \in [H_{min},H_{max}]$ client-edge aggregations before pushing the edge model to the cloud server. Taking $\eta < \frac{1}{\beta}$, the convergence upper bound of HiFL after $T_c$ global updates on the cloud server can be expressed as,
	\begin{equation}
		\small
		\label{upper_bound_convex}
		\begin{aligned}
			\mathbb{E}[F(\omega(T_{c}) - F(\omega^*)] &\le \kappa[F(\omega(0))-F(\omega^*)]\\
			&+ (1-\kappa)\frac{A_1V + A_{2}\delta_{max} + A_{3}\Delta}{B},\\ 
		\end{aligned}
	\end{equation}
	where $\kappa = (1-\alpha_{\tau} + \alpha_{\tau}(1-\eta\mu)^{cH_{min}})^{T_{c}}$, $A_{1} = \frac{1}{2\mu}$, $A_{2} = \rho H_{max}(\frac{(\eta\beta+1)^c-1}{\beta}-\eta c)$, $A_{3} = \frac{cH_{max}\eta}{2}$and $B =1-(1-\eta\mu)^{cH_{min}}$.

\begin{IEEEproof}
	Please refer to Appendix B of the separate supplementary file for details.
\end{IEEEproof} 
\end{theorem}

\begin{theorem}
\label{theorem_non_convex}
Suppose the loss function $F_k$ is $\mu$-weakly convex (which can be non-convex if $\mu > 0$), and each edge node executes $H \in [H_{min},H_{max}]$ client-edge aggregations before pushing the edge model to the cloud server. Taking $\eta < \min(\frac{1}{\beta},\frac{2}{\tilde{\mu}-\mu})$, the convergence upper bound of HiFL after $T_c$ global updates on the cloud server can be expressed as,
\begin{equation}
	\small
	\label{upper_bound_non_convex}
	%\small
	\begin{aligned}
		\mathbb{E}[F(\omega(T_{c}) - F(\omega^*)]
		&\le \tilde{\kappa}[F(\omega(0))-F(\omega^*)]\\	
		&+ (1-\tilde{\kappa})\frac{\tilde{A_1}\tilde{V}+ \tilde{A_2}\tilde{\delta}_{max} + \tilde{A_3}\tilde{\Delta}}{\tilde{B}},  
	\end{aligned}
\end{equation}
where $\tilde{\kappa} = (1-\alpha_{\tau}+\alpha_{\tau}[(1-\frac{\eta(\tilde{\mu}-\mu)}{2})^{cH_{min}}])^{T_{c}}$, $\tilde{A_1} =\frac{12\rho^{2}H_{max}}{(\tilde{\mu}-\mu)^3}+\frac{5\tilde{\mu}-\mu}{2(\tilde{\mu}-\mu)^2}$, $\tilde{A_2}= \frac{\tilde{\rho}H_{max}}{\tilde{\beta}}((\eta\tilde{\beta}+1)^c-1-\tilde{\beta}\eta c)$, $\tilde{A_3} = \frac{2H_{max}}{\tilde{\mu}-\mu}$ and $\tilde{B} = 1-(1-\frac{\eta(\tilde{\mu}-\mu)}{2})^{cH_{min}}$.
\begin{IEEEproof}
	We first give the convergence guarantee between the client model and the virtual cluster model and then provide the details of convergence analysis for the cloud model in Appendix C of the separate supplementary file.
\end{IEEEproof} 
\end{theorem}

Based on Theorem \ref{theorem_convex} and Theorem \ref{theorem_non_convex}, we draw the following two notable remarks for the convergence of HiFL.

\begin{remark}
	(Convergence rate.) The hyperparameter $\alpha_\tau$ controls the convergence rate of HiFL. Since $\alpha_\tau$ increases with the decrease of $\tau$, if a smaller $\tau$ is adopted, $\kappa$ will decrease to 0 faster as the total number of global aggregations $T_c$ grows, indicating a faster convergence rate.
\end{remark}
\begin{remark}
	(Convergence bound.) When $T_c \rightarrow \infty$, $\kappa \rightarrow 0$, the convergence bound is reduced to $\frac{A_1V + A_{2}\delta_{max} + A_{3}\Delta}{B}$ for strongly convex function, which is dominantly affected by the stochastic gradient of client $V$, the client-edge divergence $\delta_{max}$, and the edge-cloud divergence $\Delta$. Here, the values of the two coupled items, $\delta_{max}$ and $\Delta$, are determined by the client-edge association strategy. Similar observations can be found for weakly convex function.
\end{remark}
\begin{remark}
	(Impact of $\tau$ on convergence bound.) The right side of Eqn. (19) can be reformulated as $U = (C_{1})^{T_{c}}(C_2-C_3)+C_3$, where $C_1 = 1-\alpha_{\tau} + \alpha_{\tau}(1-\eta\mu)^{cH_{min}}$, $C_2 = F(\omega(0))-F(\omega^*)$ and $C_3 = \frac{A_1V + A_{2}\delta_{max} + A_{3}\Delta}{B}$. Since $C_2$ is usually very large, we assume $C_2 - C_3 > 0$. Hence, $U$ monotonically increases with $C_1$. As $C_1$ increases with $\tau$, given a fixed value of $T_c$ in practice, a bigger value of $\tau$ indicates a bigger upper bound $U$. Similar observations can be found for weakly convex function.
\end{remark}

\section{\small HiFlash: HiFL with Adaptive Staleness Control and Heterogeneity-aware Client-Edge Association}
\label{SectionHiFlash}
In this section, with the theorectical analysis above, we first conduct a preliminary evaluation on the performance of HiFL with different model staleness values and client-edge association mechanisms. Inspired by the empirical insights from experimental results, we then devise an enhanced design of HiFL, named HiFlash, with adaptive staleness control and heterogeneity-aware client-edge association to achieve high efficiency.

\begin{figure}[!t]
	\centering
	\subfigure[]{
		\includegraphics[width=0.51\linewidth]{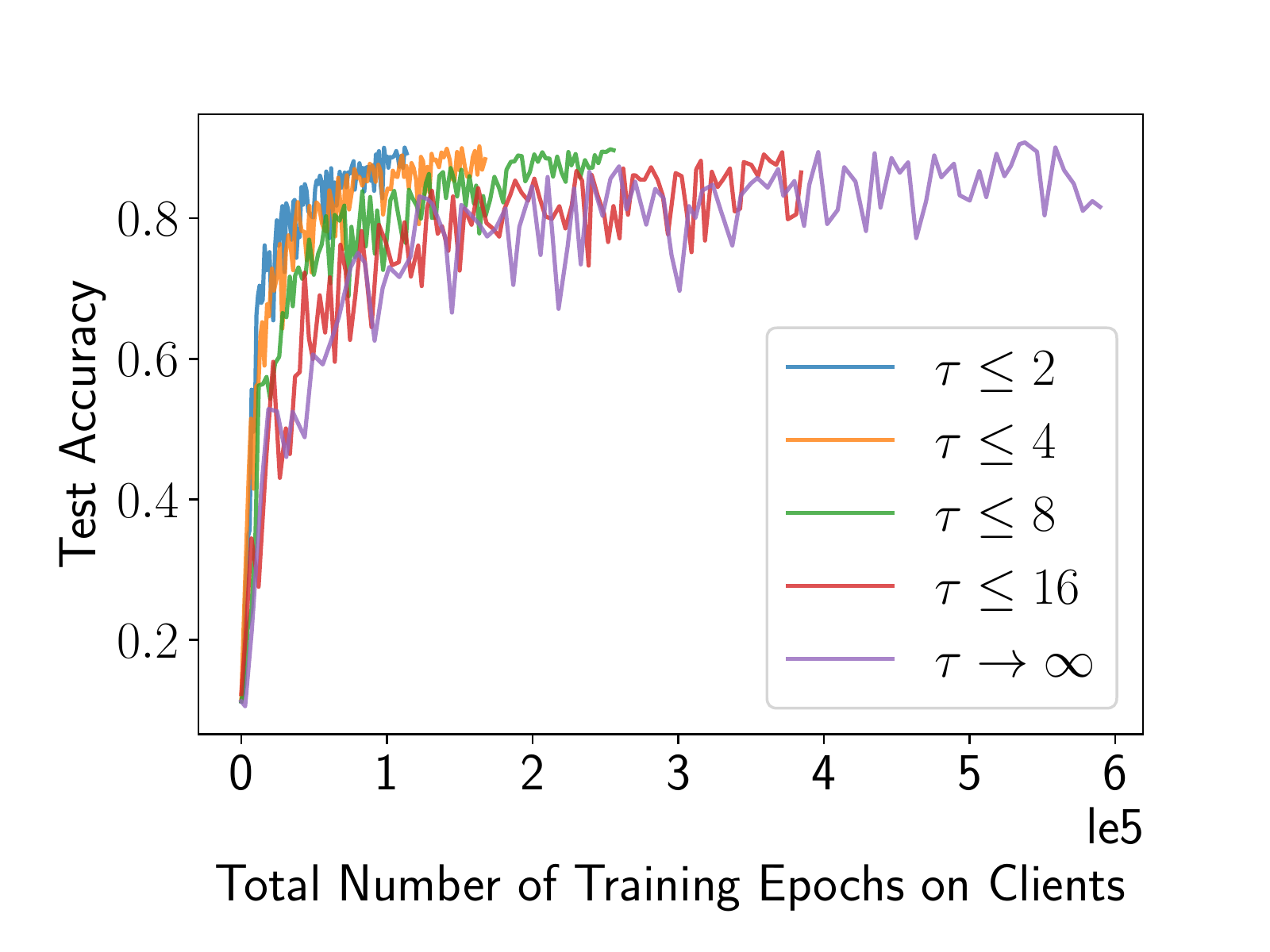}
	} 
	\subfigure[]{
		\includegraphics[width=0.43\linewidth]{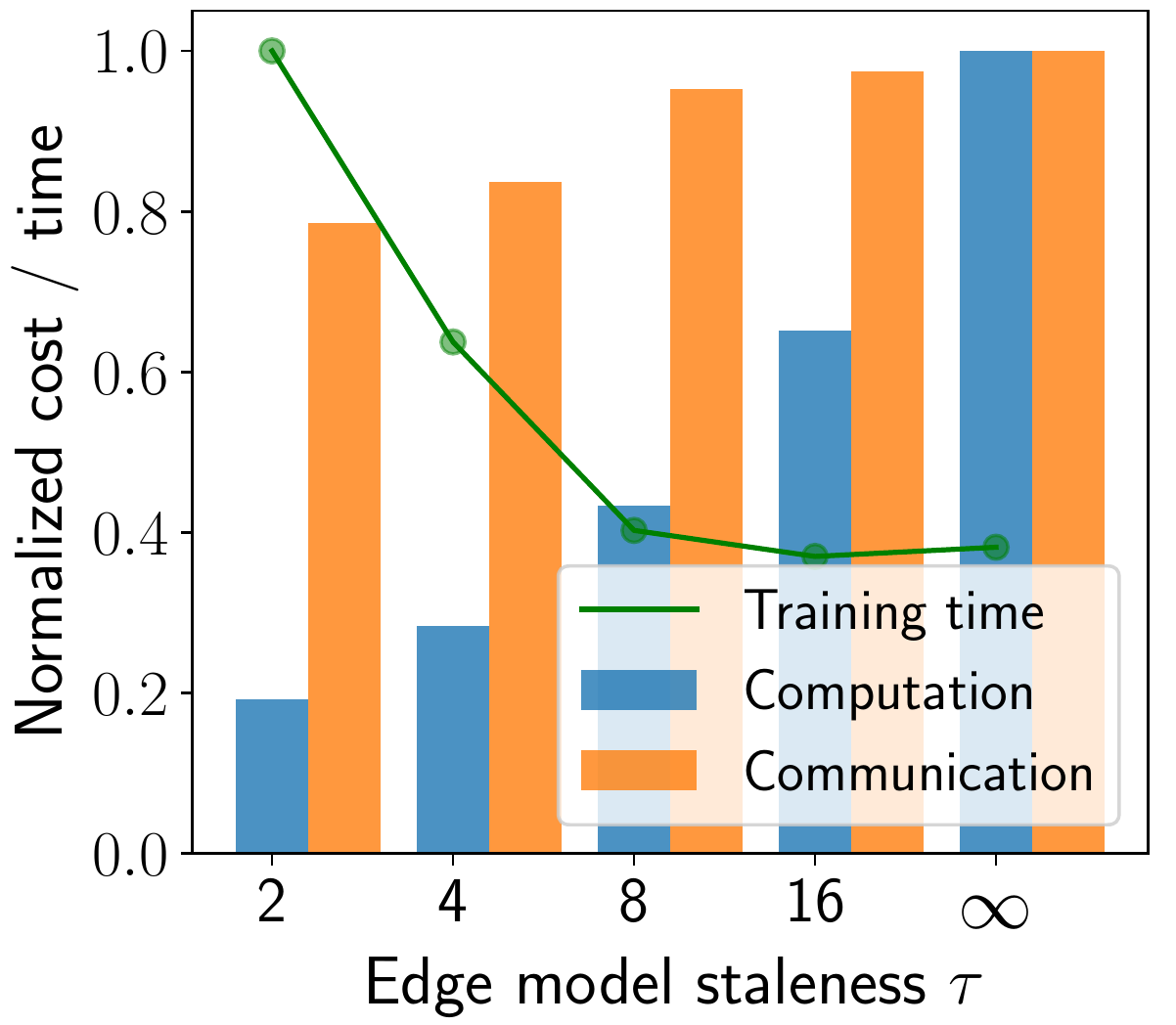}
	} 	 
	\vspace{-10pt}
	\caption{The model performance of HiFL in different aspects with varying values of edge model staleness threshold $\tau$ on MNIST dataset. (a) shows that a bigger value of $\tau$ will cause huge accuracy fluctuation. (b) gives the computation cost of clients, communication cost of edges and the training time of HiFL with varying $\tau$.}
	\label{StalenessTest}
	\vspace{-10pt}
\end{figure}

\subsection{Performance of HiFL in Deployment}
\label{ModelPerformanceofHiFL}
We first use MNIST dataset \cite{lecun1998gradient} as an example to study the model training performance of HiFL under varying model staleness values from both model performance and system cost perspectives. As depicted in Fig. \ref{StalenessTest}(a), $\tau \le 8$ means that all the edge nodes have the same maximum staleness threshold, which is 8. Hence, each edge node can upload its trained edge model with different $\tau$ to the cloud if $\tau$ is less than 8. For HiFL without staleness control (e.g., $\tau \rightarrow \infty$), all the edge models uploaded by the edge nodes can be utilized for the cloud model updating, no matter the edge model staleness. As we can see, it requires $5.3 \times 10^5$ training epochs on clients to reach a target test accuracy of 0.9 for HiFL with $\tau \rightarrow \infty$, since large model staleness results in slow convergence and dramatic accuracy fluctuation. While HiFL with staleness-restricted adjustment only allows edge model updates within a smaller staleness, ensuring a satisfactory convergence speed. For example, when $\tau \le 2$, the cloud model can reach a test accuracy of 0.9 within $9.8 \times 10^4$ training epochs on clients, less than $1/5$ of the computation cost in the case of $\tau \rightarrow \infty$.

Besides, the overall system cost should be better quantified and jointly considered in realistic large-scale FL system. Thus, we introduce three performance metrics for HiFL: \textit{total training time of the cloud}, \textit{communication cost of the edges}, and \textit{computation cost of the clients}. As observed in Fig. \ref{StalenessTest}(b), higher communication and computation costs are incurred in HiFL without staleness control in the long run. Nevertheless, smaller $\tau$ indicates decreased model parallelism, where fewer edge nodes are allowed to simultaneously train the model, considerably prolonging the training time of FL model learning. For example, in HiFL with $\tau = 0$, an edge can perform model training only when all the others are idle. Hence, staleness control for HiFL is critical for fast and cost-efficient model learning, which should be well designed to achieve a better trade-off between training time and cost efficiency.

\begin{figure}[!t]
	\centering
	\includegraphics[width=0.7\linewidth]{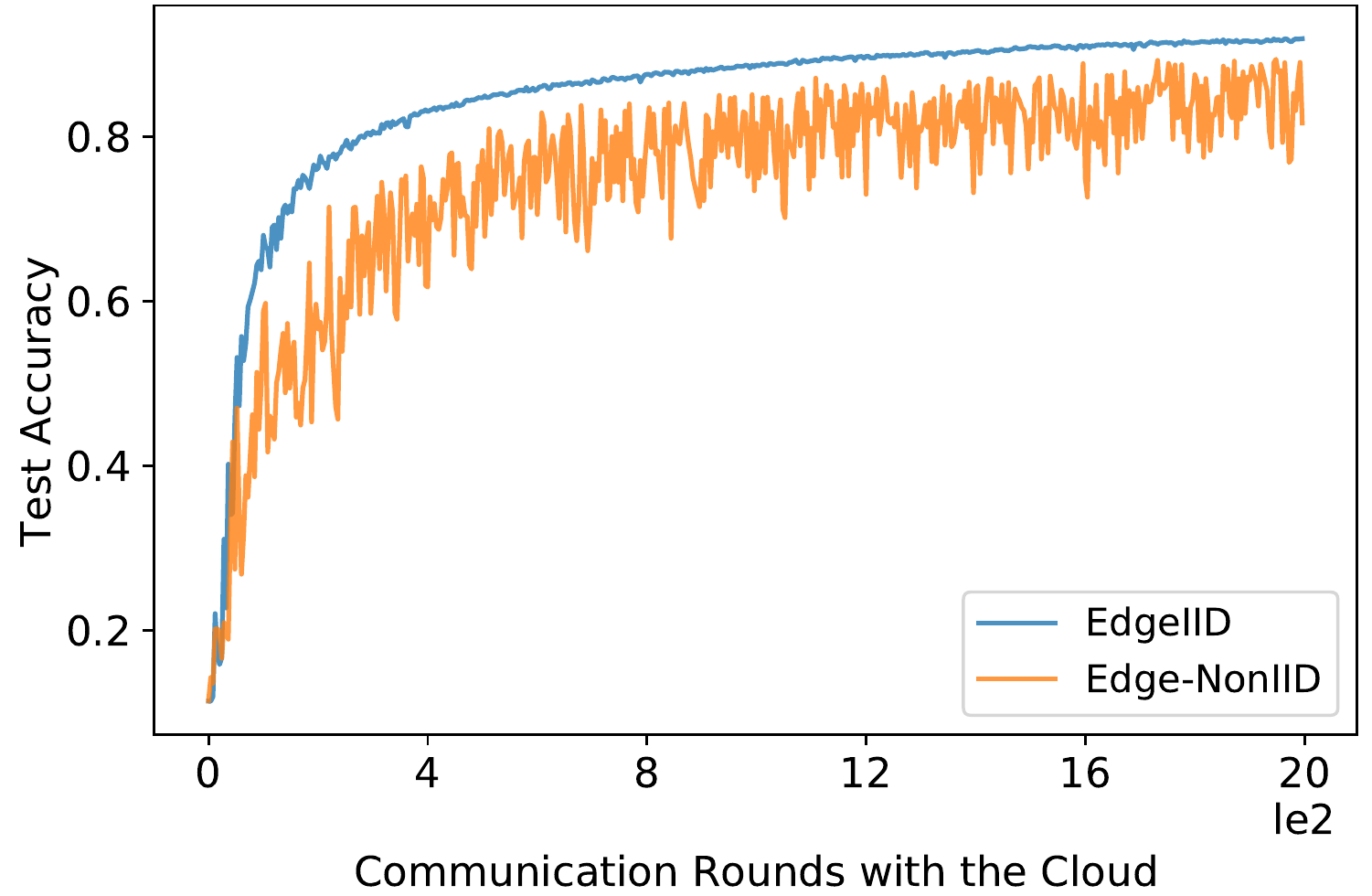}
	\vspace{-5pt}
	\caption{ EdgeIID association strategy results in smaller edge-cloud model divergence and ultimately achieves fast convergence and high model accuracy than Edge-NonIID.}
	\label{EdgeIID}
	\vspace{-10pt}
\end{figure}

Since the data heterogeneity \cite{lee2020accurate} can be a critical issue in FL, we study the influence of varying client-edge divergences and edge-cloud divergences on HiFL via multiple different client-edge association strategies. A useful insight is derived from the results, that is, edge-cloud divergence $\Delta$, as the dominant factor,  negatively impacts the cloud model accuracy. As shown in Fig. \ref{EdgeIID}, we consider a FL system with a cloud server, 10 edge nodes and 100 clients. Each client owns samples from only one single class in MNIST dataset. Edge-IID means that the clients are clustered into different edge groups and the data distributions on the edge nodes are IID (e.g., identical number of samples from 10 classes). While in Edge-NonIID case, the samples maintained by an edge node are from 5 classes. Edge-IID association strategy groups the clients with a smaller edge-cloud model divergence, and ultimately leads to fast convergence and high accuracy. Therefore, given the clients with Non-IID distributions, a heterogeneity-aware client-edge association strategy is desired to make the data distributions on the edge nodes similar to the global IID distribution. 

Motivated by the observations above, we devise HiFlash, an enhanced HiFL approach equipped with adaptive staleness control at the edge-cloud layer and heterogeneity-aware association at the client-edge layer, as elaborated below.

\subsection{Adaptive Staleness Control at Edge-Cloud Layer}
Fixed staleness control (e.g., $\tau <= 8$ in Fig. \ref{StalenessTest}) requires a predefined staleness threshold for all the edge nodes, which can work poorly in the complex dynamic FL environment (e.g., highly dynamic communication capabilities of edges, time-varying number of current training edges) and further degrades model performance. Thus, we design an adaptive edge staleness threshold for the edge nodes which are willing to join the global model training based on the condition of its control domain (e.g., the computation resources of clients, the communication capabilities and training time of edges). Specifically, we formulate the system cost model and then adopt a deep reinforcement learning approach to dynamically control the staleness threshold.

\begin{figure}[!t]
	\centering
	\includegraphics[width=0.75\linewidth]{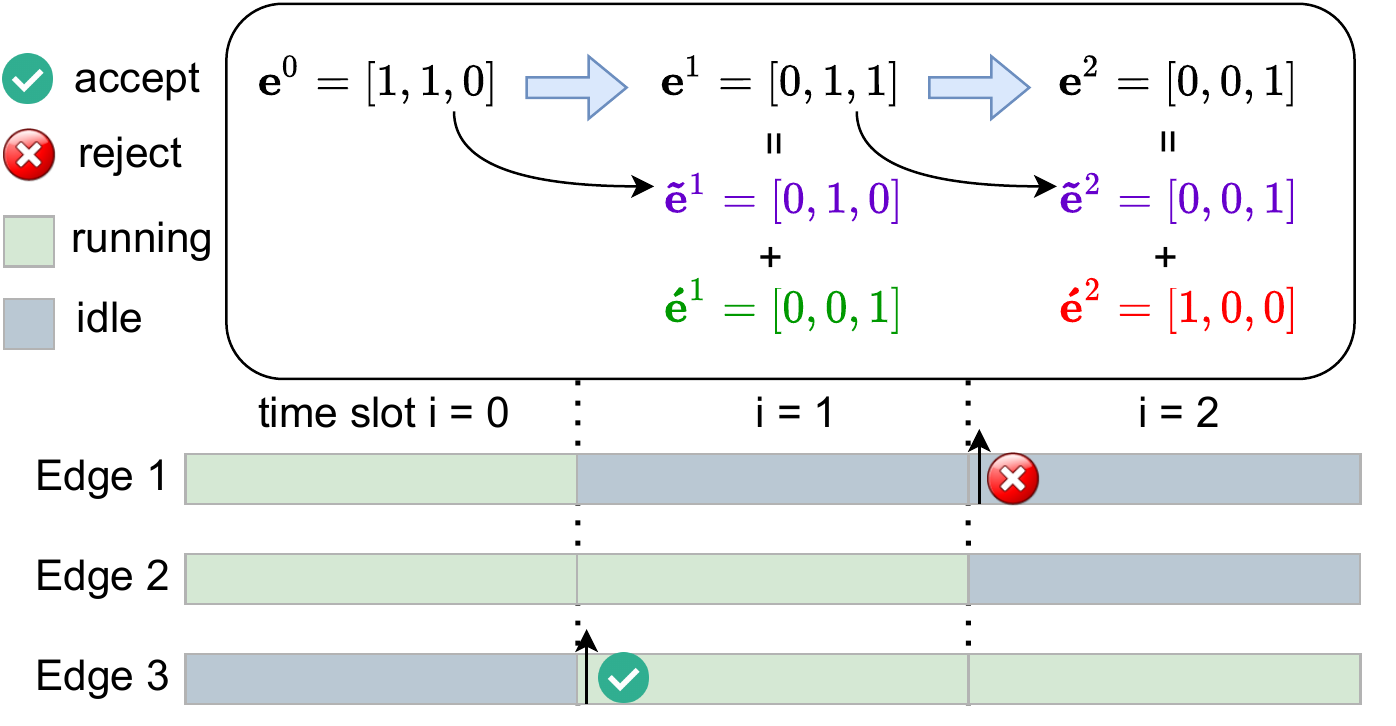}
	\vspace{-5pt}
	\caption{The illustration of the relations among $\mathbf{e}^{i}$, $\mathbf{\tilde{e}}^{i}$ and $\mathbf{\acute{e}}^{i}$.}
	\label{e_relation}
	\vspace{-10pt}
\end{figure}

\textbf{System cost model.} To fully characterize the environment dynamics, we adopt a slotted structure for staleness control to divide a long-term time horizon into a series of discrete time slot. Note that the length of each time slot is usually short, thus we assume there are at most one edge node sends a check-in request to the cloud at the beginning of time slot $i$. Similarly, at most one edge node will finish the edge model training and upload the model updates to the cloud server at the end of a time slot. We define the running/idle modes of edges at time slot i as $\mathbf{e}^{i}$, which is composed of the edges $\mathbf{\tilde{e}}^{i}$ that do not finish the edge model training task at previous time slot , and the edge $\mathbf{\acute{e}}^{i}$ which sends a check-in request and is accepted by the cloud server for participating the FL training. An example of the relationship among the definitions of $\mathbf{e}^{i}$, $\mathbf{\tilde{e}}^{i}$ and $\mathbf{\acute{e}}^{i}$ is illustrated in Fig. \ref{e_relation}.

\textit{1) Computation cost:} At a given time slot $i$, we define the computation cost of an edge node $m$ as the sum of computation cost of its associated clients:
\begin{equation}
	\label{com_cost_def1}
	\small
	C_{comp}^{i,m} = \sum_{k \in \mathcal{C}_m}C_{comp}^{i,k},
\end{equation}
where $C_{comp}^{i,k}$ denotes the computation cost of client $k$. Similarly to many existing works \cite{feng2021min}, \cite{zhan2020experience}, \cite{9445589}, following the empirical measurement study \cite{liu2020client}, we assume $C_{comp}^{i,k} = \frac{cD_k\zeta_{k}}{f_{i,k}}$, where $f_{i,k}$ is the processing speed of client $k$ at time slot $i$, and $\zeta_{k}$ is processing density for client $k$ \footnote{It is possible to train local model with GPU for devices with GPU resources, and accordingly, the computation cost is calculated with GPU cycle frequency and GPU processing density of the devices, which can be obtained by measurements.}. $D_k$ is the total number of bits for the training data of client $k$ in one local iteration and $c$ is the number of local iterations. Hence, the product of $c$ and $D_k$ indicates the workload for client $k$. The computation cost of all clients at time slot $i$ is denoted as
\begin{equation}
	\label{com_cost_def2}
	\small
	C_{comp}^{i} = \sum_{m=1}^{M} e_{m}^{i} \cdot C_{comp}^{i,m},
\end{equation}
where $e_{m}^{i} \in \{0,1\}$ indicates the idle/running modes of edge node $m$ at time slot $i$. 

\textit{2) Communication cost:} The communication cost at time slot $i$ is denoted as
\begin{equation}
	\label{comm_cost_def}
	\small
	C_{comm}^{i} = \sum_{m=1}^{M} (e_{m}^{i}-\tilde{e}_{m}^{i+1}) \cdot C_{comm}^{i,m},
\end{equation}
where $C_{comm}^{i,m} = \sum_{k \in \mathcal{C}_m}C_{comm}^{i,m,k}$ is the communication cost of edge node $m$ at time slot $i$. Following \cite{sun2022semi}, the communication cost between edge $m$ and client $k$ at time slot $i$ is calculated by $C_{comm}^{i,m,k} = \frac{\mathbb{M}_{\omega}\times 32 \text{ bits/parameter}}{B_{i,m,k} \text{log}_{2}(1+\text{SNR})}$, where $B_{i,m,k}$ is the allocated bandwidth for client $k$ by edge node $m$ at time slot $i$, $\mathbb{M}_{\omega}$ is the number of parameters of $\omega$ and \text{SNR} is set to be $17\, \text{dB}$.

\textbf{The DRL agent for adaptive staleness control.} The primary objective of model staleness control is to minimize the total system cost (including total training time of the cloud, communication cost of the edges, and computation cost of the clients) of HiFL system while achieving a target model training performance (e.g., a target accuracy $\Omega$). Due to the complicated FL learning environment, we design an experience-driven algorithm based on DRL for adaptive staleness control. We first formulate the adaptive staleness threshold optimization problem as a MDP as follows:

\textit{1) State:} At each time slot $t$, the system state is composed of three kinds of information to characterize current HiFL training environment, as elaborated below:
\begin{itemize}
	\item The information of edge training performance consists the estimated computation cost $\mathbf{\tilde{C}}_{comp}^{i} = [\tilde{C}_{comp}^{i,1}, ...,\tilde{C}_{comp}^{i,M}]$, the estimated communication cost $\mathbf{\tilde{C}}_{comm}^{i} = [\tilde{C}_{comm}^{i,1}, ...,\tilde{C}_{comm}^{i,M}]$ and the estimated time slots $\mathbf{\tilde{T}}_{train}^{i} = [\tilde{T}_{train}^{i,1}, ...,\tilde{T}_{train}^{i,M}]$ required for each edge node to complete edge model calculation.
	\item The information of current running edges is characterized as the remaining training time of current edges $\mathbf{\tilde{e}}^{i}$, denoted as $\mathbf{\tilde{T}}_{rem}^{i} = [\tilde{T}_{rem}^{i,1}, ...,\tilde{T}_{rem}^{i,M}]$, where $\tilde{T}_{rem}^{i,m} = 0$ if edge node $m$ is idle (e.g., $\tilde{e}^{i}_{m} = 0$).
	\item The information of current check-in request of the edges $\mathbf{\acute{e}}^{i}$ indicates the edge which will be informed of a staleness threshold by the DRL agent. Note that at most one check-in request from the idle edges happens at one time slot (e.g., $|\mathbf{\acute{e}}^{i}| \in \{0,1\}$). Moreover, the edge node $m$ that requests for check-in does not belong to the set of current running edges, which means $\sum_{m=1}^{M}\acute{e}^{i}_{m}\cdot\tilde{T}_{rem}^{i,m} = 0$.
\end{itemize}
In summary, the state can be represented as $\mathbf{s}^{i} = [\mathbf{\tilde{C}}_{comp}^{i},\mathbf{\tilde{C}}_{comm}^{i},\mathbf{\tilde{T}}_{train}^{i},\mathbf{\tilde{T}}_{rem}^{i},\mathbf{\acute{e}}^{i}]$. It is worthnoting that the estimated cost and training time information in the state can be profiled and collected jointly by the edge nodes and cloud server, such that the cloud server will be aware of the cost information of an edge node with a check-in request.

\textit{2) Action:} At the beginning of each time slot $i$, the DRL agent needs to decide the maximum staleness $a^{i}$ that can tolerate based on current state $\mathbf{s}^{i}$ for the idle edge node who requests for check-in ($\acute{e}^{i}_{m} = 1$). In this paper, we set an upper bound $\tau_{max}$ for staleness threshold $a^{i}$ and a lower bound $-1$ which means the check-in request is rejected by the cloud server. Here, the rejection operation can adjust the number of running edges (control the model parallelism) and hence mitigate the straggler effect. As a consequence, the decision space of action $a^{i}$ is $\{-1,0,1,...,\tau_{max}\}$. Once an action $a^{i}$ is performed, the idle/running modes of edges will be changed based on the following equation
\begin{equation}
	\label{currentRunningEdges}
	\small
	\mathbf{e}^{i} = \begin{cases}
		\mathbf{\tilde{e}}^{i} + \mathbf{\acute{e}}^{i}, & a^{i} >= 0,\\
		\mathbf{\tilde{e}}^{i},& a^{i} < 0.
	\end{cases}
\end{equation}
\textit{3) Reward:} We define the reward function $r^{i}$ as the total system cost at time slot $i$:
\begin{equation}
	\label{reward}
	\small
	r^{i}=-\sigma_{1}C_{comp}^{i}-\sigma_{2}C_{comm}^{i}-1,
\end{equation}
where $\sigma_{1}$ and $\sigma_{2}$ are two penalty factors for computation cost and communication cost, respectively.

The DRL agent aims to find the action (staleness threshold) which can minimize the long-term system cost, while guaranteeing a certain level of learning quality (e.g., reaching a target accuracy $\Omega$). Thus, the problem can be formulated to maximize the expectation of the cumulative discounted reward starting from time slot $i$ given by:
\begin{equation}
	\label{Reward}
	\small
	R^{i} = \sum_{\hat{i}=1}^{I}\gamma^{\hat{i}-1}r^{i+\hat{i}-1} = \sum_{\hat{i}=1}^{I}\gamma^{\hat{i}-1}(-\sigma_{1}C_{comp}^{i+\hat{i}-1}-\sigma_{2}C_{comm}^{i+\hat{i}-1}-1),
\end{equation}
where $\gamma \in (0,1]$ is a factor discounts future rewards and $I$ is the total training slots for reaching the target model accuracy.

We now explain the motivations of the reward design. The reward $r^{i}$ is defined as a weighted sum of computation cost, communication cost and training time. The first two terms incentivize the agent selects action that results in smaller communication and computation costs. The last term, $-1$, encourages the agent to complete training in fewer time slots (e.g., a smaller value of the total training slots $I$). By adjusting the non-negative parameters $\sigma_{1}$ and $\sigma_{2}$, it is able to meet diverse preferences of different FL learning tasks on resource efficiency and learning time. For example, we can assign higher weights to resource cost so that smaller staleness threshold are prefer to be chosen in fear of severe staleness effect and resource waste. While smaller weights of resource cost indicates that training time is critical for a FL training task and more decisions with bigger staleness threshold are made to facilitate more parallel model training.

\begin{figure}[!t]
	\centering
	\includegraphics[width=0.99\linewidth]{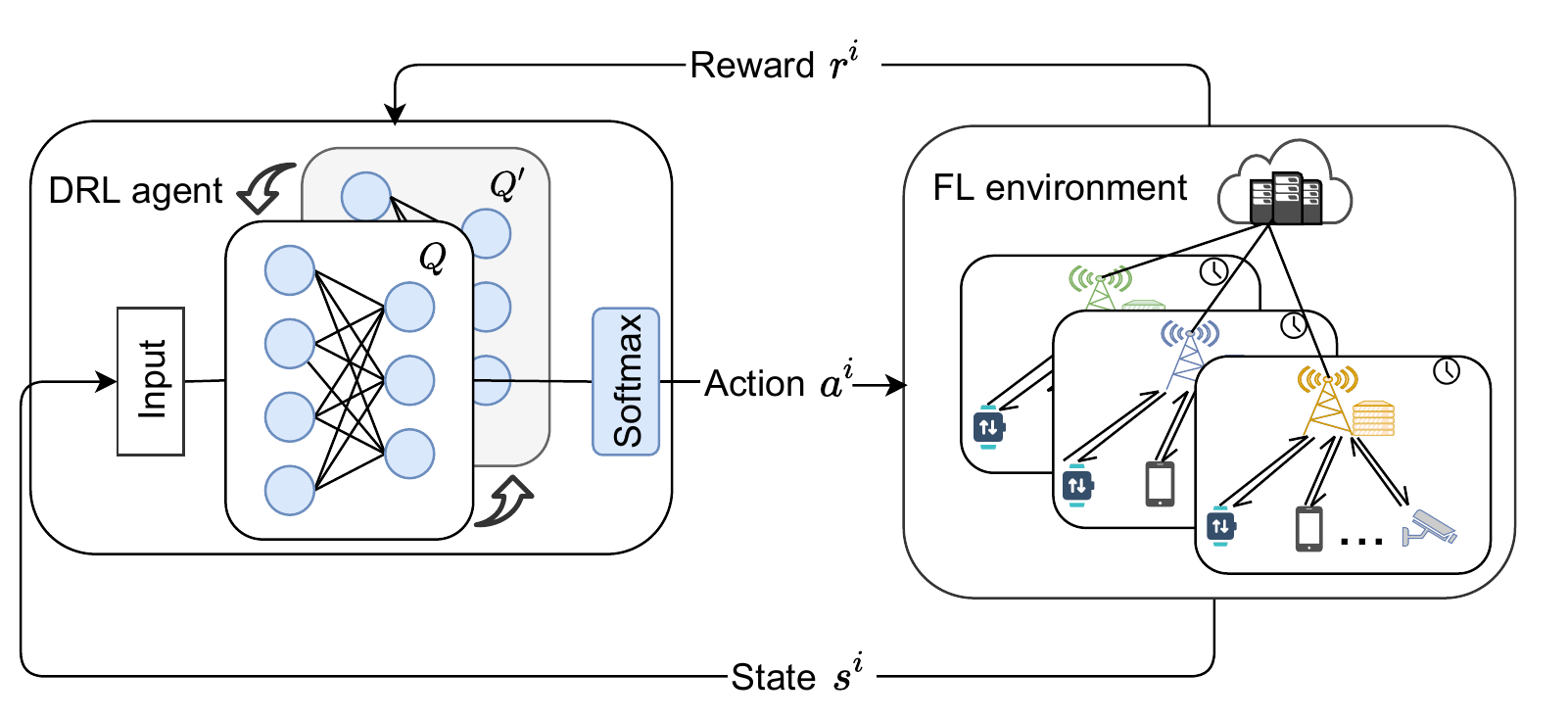}
	\vspace{-10pt}
	\caption{The DRL agent interacting with the FL environment.}
	\label{DQN}
	\vspace{-10pt}
\end{figure}

\textbf{Training procedure of DRL agent.}
Considering the continuous and high-dimensional space and the limited available traces from FL tasks, we adopt DQN as the DRL agent to efficiently learn the  optimal staleness control policy. Due to the complicated dynamics in FL system, overestimation issue can easily arise due to insufficient exploration by DQN. Hence, to solve the overestimation problem, we propose to use Double Deep Q-learning (DDQN) to learn the approximator $Q_{\pi}(\mathbf{s}^{i},a^{i})$ that approximates to the optimal action-value function $Q^{*}(\mathbf{s}^{i},a^{i})$ \cite{van2016deep}. DDQN introduces a double estimator $Q^{\prime}(\mathbf{s}^{i},a^{i},\theta^{\prime}_{i})$, which frozes every $U$ updates, to stabilize the action-value function estimation. 

To train the DRL agent, as depicted in Fig. \ref{DQN}, current state information $\mathbf{s}^{i}$ is fed into action-value function $Q$ and then DQN generates action $a^{i}$ as the staleness threshold for the edge node which is willing to join the FL model training. After interacting with the FL environment for several rounds, the DRL agent samples a few state-action pairs from the experience memory to solve Eqn. (\ref{DQN_objective}) as 
\begin{equation}
	\small
	\label{DDQN_objective}
	\arg \min_{\theta_{i}} \mathbb{L}(\theta_{i})=(Y^{\prime}_{i} - Q_{\pi}(\mathbf{s}^{i},a^{i},\theta_{i}))^2,
\end{equation}
where the target $Y^{\prime}_{i}$ is defined as
\begin{equation}
	\small
	\label{Y_target}
	Y^{\prime}_{i}=r^{i} + \gamma  Q^{\prime}(\mathbf{s}^{i+1},\arg\max_{a^{i+1}}Q(\mathbf{s}^{i+1},a^{i+1},\theta_{i}),\theta^{\prime}_{i}),
\end{equation}
where $\theta_{i}$ is the online parameters updated per time step and $\theta^{\prime}_{i}$ is the parameters of double estimator $Q^{\prime}$. 

The action-value function $Q(\mathbf{s}^{i},a^{i})$ is updated by minimizing $\mathbb{L}(\theta_{i})$ with gradient descent, i.e.,
\begin{equation}
	\small
	\label{L_update}
	\theta_{i+1} = \theta_i + \alpha_{Q}(Y^{\prime}_{i}-Q(\mathbf{s}^{i},a^{i},\theta_{i})\nabla_{\theta_{i}} Q(\mathbf{s}^{i},a^{i},\theta_{i})),
\end{equation}
where $\alpha_{Q}$ is a scalar step size. Besides, the classical $\epsilon$-greedy policy is adopted in DDQN model training to aid exploration \cite{van2016deep}. The details of the DRL-based staleness control process is elaborated in Algorithm \ref{DDQN_training}.

\begin{algorithm}[htb]
	\caption{DDQN training procedure for adaptive staleness control}
	\label{DDQN_training}
	\small
	\begin{algorithmic}[1]
		\Require discount factor $\gamma$, target accuracy $Acc_{target}$, experience memory maximum size $B_{max}$, update frequency of target network $U$.
		\State Initialize experience memory $\mathbf{E}_{exp}$
		\State Initialize action-value network $Q$ with initial weight $\theta$
		\State Initialize the target network $Q^{\prime}$ as a copy of $\theta$
		\State \textbf{for} each episode $epi = 1,2,3,...$ \textbf{do}
		\State \quad \textbf{for} $i = 1,2,...,I$ \textbf{do}
		\State \quad \quad Get current state $\mathbf{s}^{i}$ from the FL learning environment
		\State \quad \quad Generate action $a^{i}$ based on $\epsilon$-greedy policy and execute it
		\State \quad \quad Observe reward $r^{i}$ and next state $\mathbf{\tilde{s}}^{i}$
		\State \quad \quad Store the tuple $(\mathbf{s}^{i},a^{i},r^{i},\mathbf{\tilde{s}}^{i})$ into $\mathbf{E}_{exp}$
		\State \quad \quad \textbf{if} $|\mathbf{E}_{exp}| > B_{max}$ \textbf{then}
		\State \quad \quad \quad Remove the oldest tuple
		\State \quad \quad \textbf{end if}
		\State \quad \quad Sample a minibatch of tuples $(\mathbf{s},a,r,\mathbf{\tilde{s}})$ from $\mathbf{E}_{exp}$
		\State \quad \quad Get target values with Eqn. (\ref{Y_target})
		\State \quad \quad Train and update the action-value network with the objective of Eqn. (\ref{DDQN_objective}) and Eqn. (\ref{L_update})
		\State \quad \quad Update the target network as a copy of the weights of action-value network every $U$ steps
		\State \quad \quad Get the accuracy $Acc$ of the cloud model
		\State \quad \quad \textbf{if} $Acc >= Acc_{target}$ \textbf{then}
		\State \quad \quad \quad break out of current episode
		\State \quad \quad \textbf{end if}
		\State \quad \textbf{end for}
		\State \textbf{end for}
	\end{algorithmic}
	\vspace{-2pt}
\end{algorithm}
The DRL agent is deployed in the scheduler component of the cloud server, thus the scheduler can inform an edge about the staleness threshold while distributing the latest global model to the edge who sends a check-in request.

\subsection{Heterogeneity-aware Association at Client-Edge Layer} Inspired by the convergence analysis in Section \ref{SectionConvergenceAnalysis} and the discussion in Section \ref{ModelPerformanceofHiFL}, we identify the controllable factors for learning performance enhancement and aim to design a client-edge association mechanism that minimizes the edge-cloud model divergence. Due to the synchronous mechanism, the edge model aggregation can be conducted until the associated slowest client uploads its newly-updated model. Hence, the resource heterogeneity (e.g., response latency) among the clients should also be taken into consideration. We strike a nice balance between data heterogeneity of edges and resource heterogeneity inherent in clients to establish a heterogeneity-aware client-edge association mechanism for fast and accurate cloud model learning. 

For data heterogeneity in FL, we primarily investigate the label distribution skew which always exists in real-world applications \cite{zhang2022federated}. As the edge-cloud model divergence is attributed to data heterogeneity between the edge node and the cloud server, we resort to Jensen-Shannon (JS) divergence \cite{majtey2005jensen}, which is based on Kullback-Leibler (KL) divergence, to calculate the dissimilarity between two datasets. Considering two probability distributions $P$ and $P^{\prime}$, the JS divergence between $P$ and $P^{\prime}$ is defined as
\begin{equation}
	\label{JS}
	\footnotesize
	\begin{gathered}
		JS(P||P^{\prime}) = \frac{1}{2}KL(P||\frac{P+P^{\prime}}{2})+\frac{1}{2}KL(P^{\prime}||\frac{P+P^{\prime}}{2}),\\
		KL(P_{1}||P_{2}) = \sum_{x \in X}P_{1}(x)log\frac{P_1(x)}{P_2(x)}.
	\end{gathered}
\end{equation}

JS divergence has appealing properties of symmetry and normalized values between $[0,1]$, which is contrast with unbounded KL divergence. $JS(P||P^{\prime}) = 0$ indicates identical distributions of $P$ and $P^{\prime}$ while $JS(P||P^{\prime}) \rightarrow 1$ means the distributions are considered highly distant. When facing feature distribution skew, we can leverage FedBN \cite{li2020fedbn} as the aggregation method to help harmonizing local feature distributions in the collaborative training process, which would be an interesting future research direction.

For resource heterogeneity of clients, we first define the response latency for client $k$ associated with edge node $m$ as:
\begin{equation}
	\small
	L^{m,k} = l_{comp}^{k} + l_{comm}^{m,k},
\end{equation} 
where $l_{comp}^{k}$ is the average value of the computation latency $C_{comp}^{i,k}$ for client $k$  over the time span and $l_{comm}^{m,k}$ is the average value of the communication latency $C_{comm}^{i,m,k}$ between edge $m$ and client $k$ over the time span. Here, both  $l_{comp}^{k}$ and $l_{comm}^{m,k}$ can be obtained from the historical records. As a result, the latency for edge model aggregation can be formulated as 
\begin{equation}
	\small
	L^m = \max \limits_{k \in \mathcal{C}^m}L^{m,k},
\end{equation}
indicating that one edge node should form its client cluster by selecting clients with lower response latency to mitigate straggler effect and accelerate model training process.

Before performing client-edge association, each of the edge nodes first probes the clients in its communication range to measure the response latency and collect the data distributions of the clients. The data distribution of a client records the proportion of samples for different classes. For instance, given an application with 4 distinct labels and a client dataset $\mathcal{D}_k$ that has one example with label 0, and two examples with label 2, the client's label distribution can be defined as $\textit{LD}_{k}(\mathcal{D}_k) = [\frac{1}{3},0,\frac{2}{3},0]$. Note that label information of a client is only revealed in an aggregated format (see Eqn. (\ref{tradeOff})) by using secure multiparty computation (e.g., privacy-preserving k-secure sum protocol \cite{sheikh2009privacy})  and after noise is added, so no violation of individual label privacy happens.

After obtaining the response latency and label distribution of clients in the communication range, the edge node $m$ can strike a balance between measured latency and edge-cloud model divergence by calculating the cost defined as \begin{equation}
	\label{tradeOff}
	\footnotesize
	COST_{k}^{m} = L^{m,k} + \lambda JS(\frac{|\mathcal{D}^m| \cdot LD^{m} + |\mathcal{D}_k|\cdot LD_{k}}{|\mathcal{D}^m| + |\mathcal{D}_k|},LD_{IID}),
\end{equation}
where $\lambda$ is a weighting parameter for the trade-off between resource heterogeneity and data heterogeneity. $\textit{LD}^{m}$ represents the label distribution on edge node $m$, the weighted average of the label distributions in current client cluster $\mathcal{C}^{m}$. $\textit{LD}_{IID}$ denotes the IID label distribution hold by the cloud server, Here, we consider the global data distribution is IID as the cloud server coordinates the model learning on multiple edge nodes, reaching a large amount of samples from different classes.

Hence, the edge node can conduct heterogeneity-aware client-edge association in the following two-way selection manner:
\begin{enumerate}
	\item  If the client cluster for edge node $m$ is empty, the edge node selects the client with the lowest response latency from the unassociated clients in its communication range. Otherwise, the edge node will select client $k$ with smallest cost $COST_{k}^{m}$
	\item If one client is currently selected by multiple edge nodes, the client will choose an edge node randomly. 	
\end{enumerate}
The two-way selection procedure continues until all the clients are associated with one edge node. The detailed heterogeneity-aware client-edge association strategy is presented in Algorithm \ref{CEA}.

\begin{algorithm}[htb]
	\small
	\caption{Heterogeneity-aware client-edge association}
	\label{CEA}
	\begin{algorithmic}[1]
		\Require Label distribution of all the clients $\{\textit{LD}_k\}_{k=1}^{N}$, the response latency $\{l_{k}^{m}\}_{k=1}^{N}$ for edge node $m \in \{1,2,..., M\}$.
		\State Construct a client pool $\textit{ClientPool} = \{1, 2,..., N\}$
		\State Construct empty client cluster for each edge node: $\mathcal{C}^m = \emptyset$ for $m \in \{1,2,..., M\}$, and empty edge set $S_{k} = \emptyset$ for each client $k \in \{1,2,...,N\}$
		\State \textbf{While} $\textit{ClientPool} \neq \emptyset$ \textbf{do}
		\State \quad \textbf{for} $m \in \{1,2,..., M\}$ \textbf{do}
		\State \quad \quad \textbf{if} $\mathcal{C}^m = \emptyset$ \textbf{then}
		\State  \quad \quad \quad Choose client $k$ with lowest response latency and set $\mathcal{C}^m =\{k\}$
		\State \quad \quad \quad Remove client $k$ from $\textit{ClientPool}$
		\State \quad \quad \textbf{else}
		\State \quad \quad \quad Choose client $k$ by minimizing Eqn. (\ref{tradeOff})
		\State \quad \quad \quad $S_{k} = S_{k} \cup \{m\}$
		\State \quad \quad \textbf{end if}
		\State \quad \textbf{end for}
		\State \quad \textbf{for} $k \in \{1,2,...,N\}$ \textbf{do}
		\State \quad \quad \textbf{if} $S_{k} \neq \emptyset$ \textbf{then}
		\State \quad \quad \quad Randomly select an edge node $m$ from $S_{k}$
		\State \quad \quad \quad $\mathcal{C}^m =\mathcal{C}^m \cup \{k\}$
		\State \quad \quad \quad Remove client $k$ from $\textit{ClientPool}$ and clear $S_{k}$
		\State \quad \quad \textbf{end if}
		\State \quad \textbf{end for}
		\Ensure Client clusters $\mathcal{C}^m$ for $m \in \{1,2,...,M\}$.
	\end{algorithmic}
	\vspace{-2pt}
\end{algorithm}

\begin{figure}[!t]
	\centering
	\includegraphics[width=0.7\linewidth]{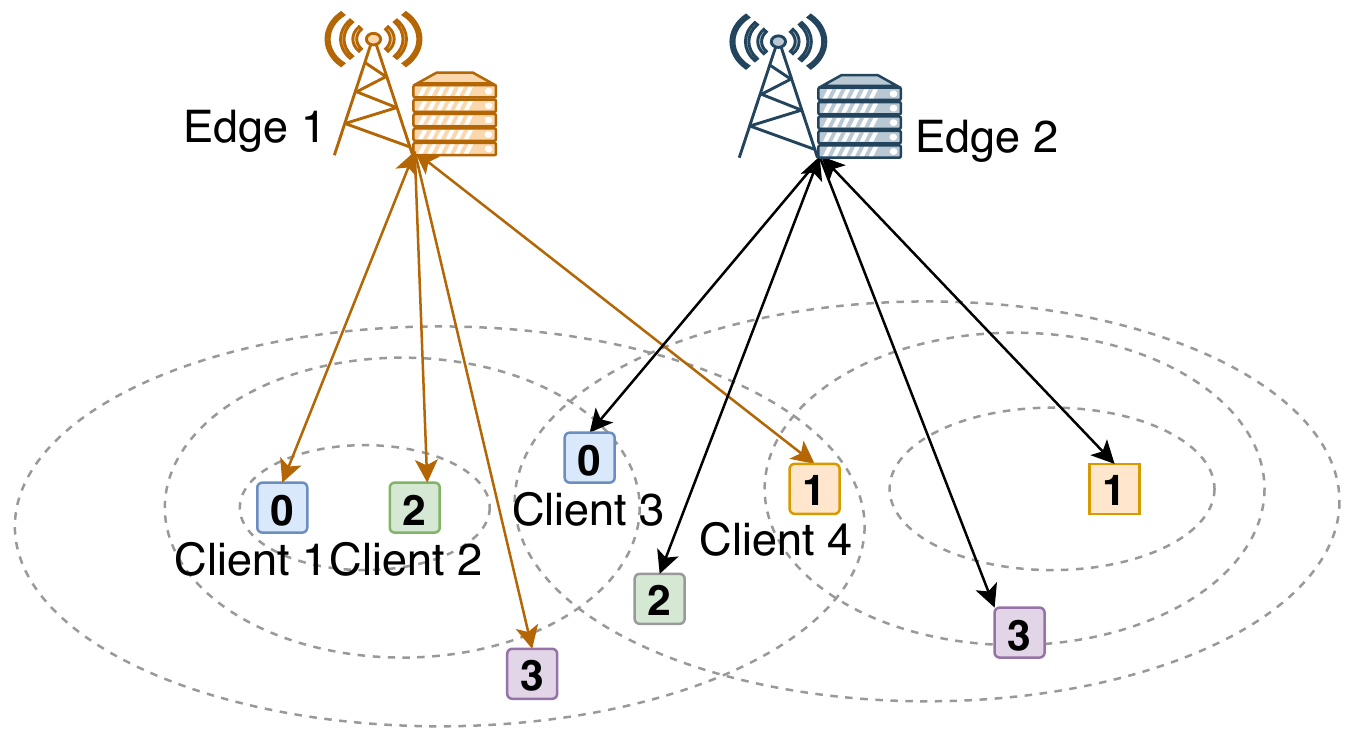}
	\vspace{-10pt}
	\caption{The contour plot of response latency of the clients communicated to different edge nodes. The samples in each client are from only one class, represented by the number in the square. For edge 1, after selecting client 1 and client 2, it is better to choose client 4 rather than client 3, taking consideration of the trade-off between response latency and data distribution of the client.}
	\label{CEA_plot}
	%\vspace{-15pt}
\end{figure}

For a more intuitive illustration, as depicted in Fig. \ref{CEA_plot}, after adding client 1 and client 2 into the client cluster of edge 1, it is better to choose client 4, rather than client 3, in order to make the data distribution of edge 1 close to the IID distribution. Similarly to many existing studies such as \cite{nishio2019client}, \cite{damaskinos2022fleet}, we consider that the clients are stationary or their locations change slowly during FL training process. This can be mainly motivated by that since FL requires intensive computing and frequent communication, most clients would like to participate FL when they are in stable conditions (e.g., when charging their batteries at home/office) \cite{kairouz2021advances}. For the case that clients' locations change fast, we can apply the adaptive client-edge association strategies to periodically update the clients' edge node selections. And some online optimization algorithms can be further leveraged to improve the performance of dynamic client-edge associations. Nevertheless, the theoretical analysis of such case is much more involved, and will be considered as a future work due to space limit.

In practice, with the label distributions of all the clients and the measured latency among clients and edges, we can easily get different client-edge association strategies under different $\lambda$. Hence, we can obtain the total JS divergence of all the edges under different $\lambda$ and then select the client-edge association strategy with smaller JS divergence for fast FL model training. HiFlash can be seen as an enhanced HiFL approach equipped with adaptive staleness control and heterogeneity-aware client-edge association, hence we can choose either HiFL or HiFlash for efficient FL model training. In HiFlash, adaptive staleness control and heterogeneity-aware client-edge association are designed to alleviate the staleness issue in asynchronous aggregation at the cloud server and data heterogeneity among the edges, respectively. As a result, it is possible to combine each of these two parts with existing works with asynchronous FL for performance enhancement, by following the similar ideas developed in our paper.

\section{Experiments}
\label{SectionExperiments}
\subsection{Simulation Settings}
In order to gauge the effectiveness of our proposed algorithm, we conduct extensive evaluations in a simulated environment with 100 clients, 10 edge nodes and a cloud server. We consider image classification as the FL task and evaluate the performance of HiFL and HiFlash with three real-world datasets: MNIST \cite{lecun1998gradient}, CIFAR10 \cite{krizhevsky2009learning} and FEMNIST \cite{caldas2018leaf}. As FEMNIST is a federated version of Extended MNIST dataset \cite{cohen2017emnist} whith 805,263 samples from 3,550 writers, we randomly select 100 writers as the clients to participate the model training of in our experiments. For the 10-class hand-written digit classification dataset MNIST, we use LeNet \cite{lecun1990handwritten} as the model trained on the clients. For the CIFAR10 dataset, a standard ResNet-18 \cite{he2016deep} model is adopted. For the 62-class hand-written digit classification dataset FEMNIST, we design a convolutional neural network (CNN) with 214,590 learning parameters as the learning model. The datasets and the corresponding models are summarized in Table \ref{dataset_model}. All the experiments are conducted on one Tesla P100 12GB GPU and the algorithms are implemented by Pytorch version 1.10.0. 

\begin{table}[!t]
	%\normalsize
	\caption{Datasets and the corresponding models. }
	\vspace{-5pt}
	\newcommand{\tabincell}[2]{\begin{tabular}{@{}#1@{}}#2\end{tabular}}
	\label{dataset_model}       % Give a unique label
	\centering
	\begin{tabular}{|c|c|c|}
		\hline
		Dataset &  Model & Parameter number\\
		\hline MNIST & LeNet & 21,840\\
		\hline CIFAR10 & ResNet-18 & 3,504,554\\
		\hline FEMNIST & CNN & 214,590\\
		\hline
	\end{tabular}
	\vspace{-10pt}
\end{table}

For the local computation of the training on each client, we employ mini-batch Stochastic Gradient Descent (SGD) with a batch size of 60 for MNIST and FEMNIST, and 50 for CIFAR10, respectively. The initial learning rates are 0.01 for MNIST and FEMNIST, and 0.1 for CIFAR10 as in \cite{liu2020client}, both of which decay exponentially at a rate of 0.99 every 100 epochs. The number of local updates $c$ for each client in one client-edge communication round is set to be 3 and the number of client-edge aggregations before pushing the edge model to the cloud server is set as $H \in \{1,2,3\}$.\footnote{The values of $c$ and $H$ depend on the computation budgets of the devices in practice. Due to the computing resource limitation of our research lab, we set small values for both $c$ and $H$, but it is sufficient to evaluate the effectiveness of HiFlash.} The hyperparameters $\alpha$ and $\upsilon$ in coefficient $\alpha_{\tau}$ can be determined by grid search in practice.

For DRL training, we set different threshold bounds ($\tau_{max}=16$ for MNIST and FEMNIST datasets, and $\tau_{max}=4$ for CIFAR10 dataset) for the three datasets as a bigger staleness threshold will result in much longer training time for the complicated CIFAR10 dataset. Hence, according to different action size, the DDQN model in the DRL agent, which is implemented by two two-layer multi-layer perceptron (MLP)  networks, has 4,607 and 4,235 trainable parameters for MNIST (and FEMNIST), and CIFAR10 datasets, respectively. The output of the MLP network passing through a softmax layer becomes the probability of selecting a staleness threshold. The DDQN is lightweighted and each training iteration takes seconds on GPU.

\textbf{Data Heterogeneity.}
For MNIST and CIFAR10 datasets, to simulate the data heterogeneity of clients in real world, we generate three kinds of data distributions for clients as below:
\begin{itemize}
	\item \textbf{IID:} Each client is randomly assigned a uniform data distribution over 10 classes.
	\item \textbf{Non-IID(1):} Each client possesses only one random class of images.
	\item \textbf{Non-IID(2):} The samples in each client are assigned from two randomly selected classes.
\end{itemize}
While the FEMNIST dataset naturally falls in the following three data heterogeneity cases:
\begin{itemize}
	\item \textbf{Label distribution skew:} The label distributions are totally different among the writers.
	\item \textbf{Feature distribution skew:} There is a natural feature distribution skew among different writers due to their different character features (e.g., stroke width, slant).
	\item \textbf{Quantity skew:} The samples in each client are ranging from $[4,525]$.
\end{itemize}

The data distribution on an edge node can be obtained by calculating the weighted average of the data distributions of its associated clients. We can use JS divergence to measure the data heterogeneity of the edge nodes.

\textbf{Resource Heterogeneity.} 
The highly heterogeneous hardware resources (CPU, network connection) among clients can be reflected by the computing latency $C_{comp}^{i,k}$ and communication latency $C_{comm}^{i,m,k}$. For the computation ability of each client $k$, we assume  $f_k \in [1, 2]$  \text{GHz} as the CPU cycle frequency and  $\zeta_{k} = 20$ \text{cycles/bit} as the number of CPU cycles to execute one bit. For the communication ability, the bandwidth $B_{k,m}$ is ranging from $1 \text{ MHz}$ to $10 \text{ MHz}$ when associated with different edge node $m$. 

\subsection{Metrics and Baselines}
\textbf{Performance Metrics.} We consider \textit{test accuracy}, \textit{the number of communications with the cloud}, and \textit{overall system cost} as three metrics for performance evaluation of HiFlash. Besides, we also calculate the \textit{response latency} $L^{m}$ and \textit{waiting time} $L^{m}_{waiting}$ during synchronous client-edge aggregation to evaluate the effectiveness of client-edge association strategy. Here, the waiting time $L^{m}_{waiting}$ is measured by the average waiting time for the straggler in the clients associated to edge node $m$, denoted as 
\begin{equation}
	\small
	L^{m}_{waiting} = \frac{\sum_{k \in \mathcal{C}^{m}}(L^{m,k}-\min \limits_{k \in \mathcal{C}^m}L^{m,k})}{|\mathcal{C}^{m}|}.
\end{equation}

\textbf{Baselines.} We compare our proposed algorithm with both traditional centralized method and federated learning based schemes for performance evaluation:
\begin{itemize}
	\item \textbf{Centralized Learning:} This scheme collects all the raw data to the cloud for training and provides an upper bound for model accuracy.
	\item \textbf{FedAvg  \cite{mcmahan2017communication}:} A cloud-based FL scheme with synchronous aggregation. Each time, it randomly selects 10 clients for local model training and global model aggregation. 
	\item \textbf{FedAsync \cite{Xie2019Asyn}:} A cloud-based asynchronous federated learning algorithm which updates the global model without waiting for straggling clients.
	\item \textbf{HierFAVG \cite{liu2020client}:} A cloud-edge-client hierarchical FL scheme that performs synchronous update in both client-edge aggregation and edge-cloud aggregation. For fair comparison, 5 clients are randomly selected for edge model aggregation and 2 edge nodes contribute to the cloud model update in each global training round. 
	\item \textbf{FedAT \cite{chai2020fedat}:} A hierarchical FL scheme that combines synchronous intra-tier training and asynchronous cross-tier training. FedAT conducts client clustering based on their response latencies, without considering the data heterogeneity of the clients. 
\end{itemize}
It is worthnoting that we adopt random staleness control mechanism with a fixed staleness threshold for the asynchronous FL schemes (e.g., FedAsync, FedAT and HiFL) for fair comparison with HiFlash. For MNIST dataset, we set a bigger value ($\tau_{max} = 16$)  to facilitate more parallel model training and shorten the total training time. While for the complex CIFAR10 dataset, we set a smaller value ($\tau_{max} = 4$) to reduce resource waste as the local training of CIFAR10 dataset incurs high resource cost. When facing new datasets, we can determine the staleness threshold based on our preferences (e.g., cost efficiency, training time) of the FL learning task.

\begin{figure}[!t]
	\centering
	\includegraphics[width=0.85\linewidth]{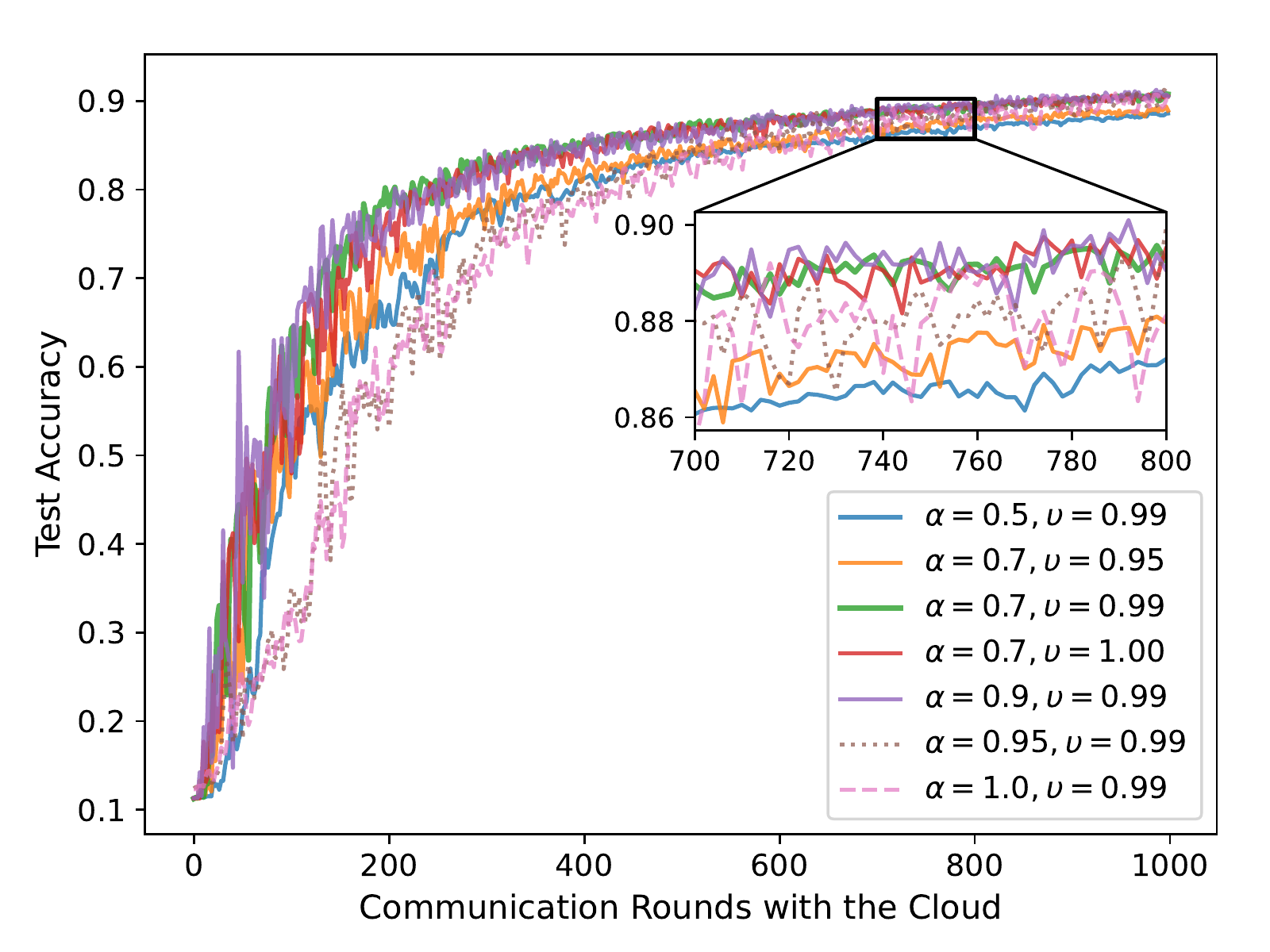}
	\vspace{-10pt}
	\caption{Test accuracy of MNIST dataset under Non-IID(2) data distribution with different mixing hyperparameters.}
	\label{alpha_tau}
	\vspace{-10pt}
\end{figure}

\begin{figure*}[!t]
	\centering
	\subfigure[Non-IID(1)]{
		\includegraphics[width=0.315\linewidth]{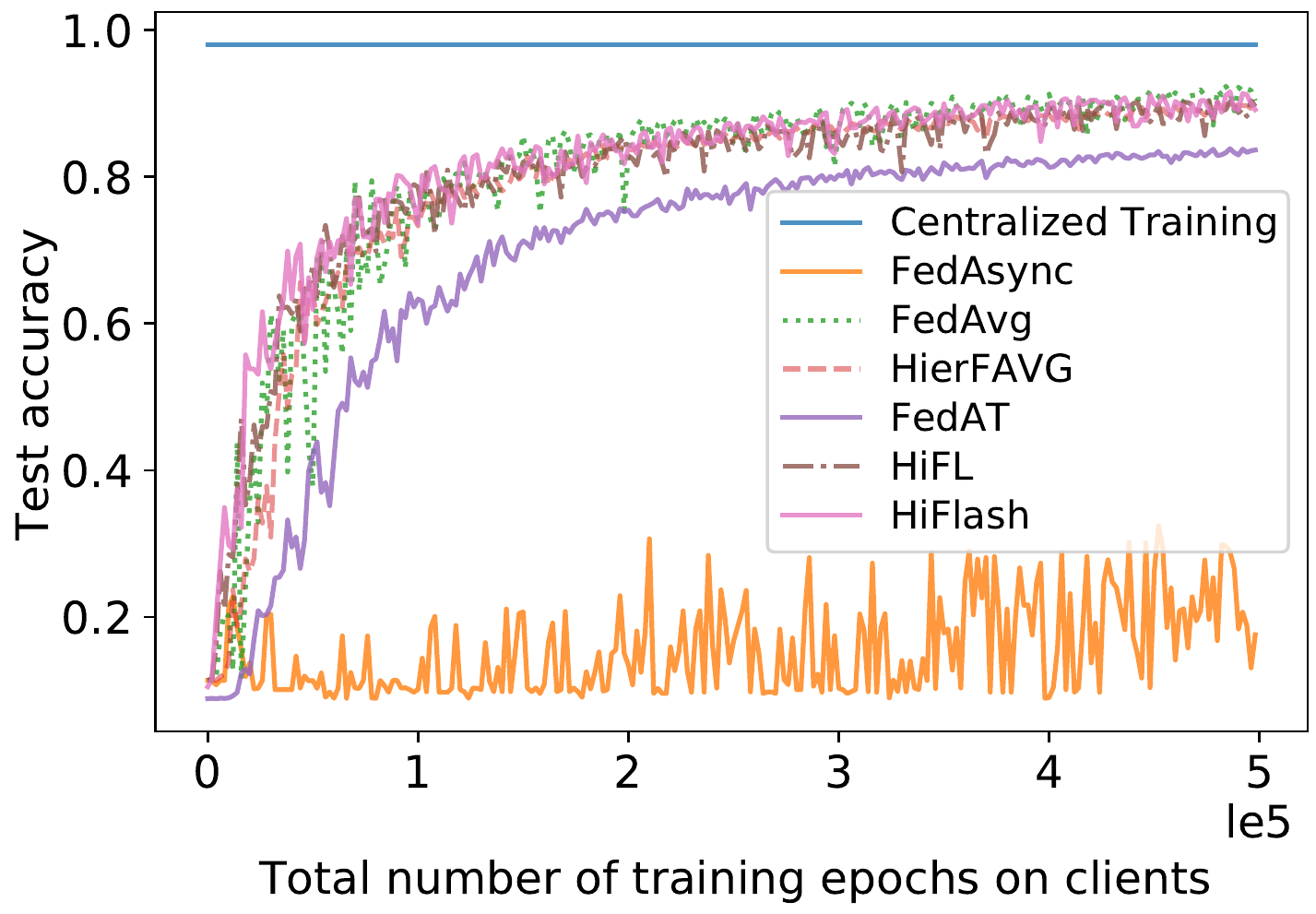}
	} 
	\subfigure[Non-IID(2)]{
		\includegraphics[width=0.315\linewidth]{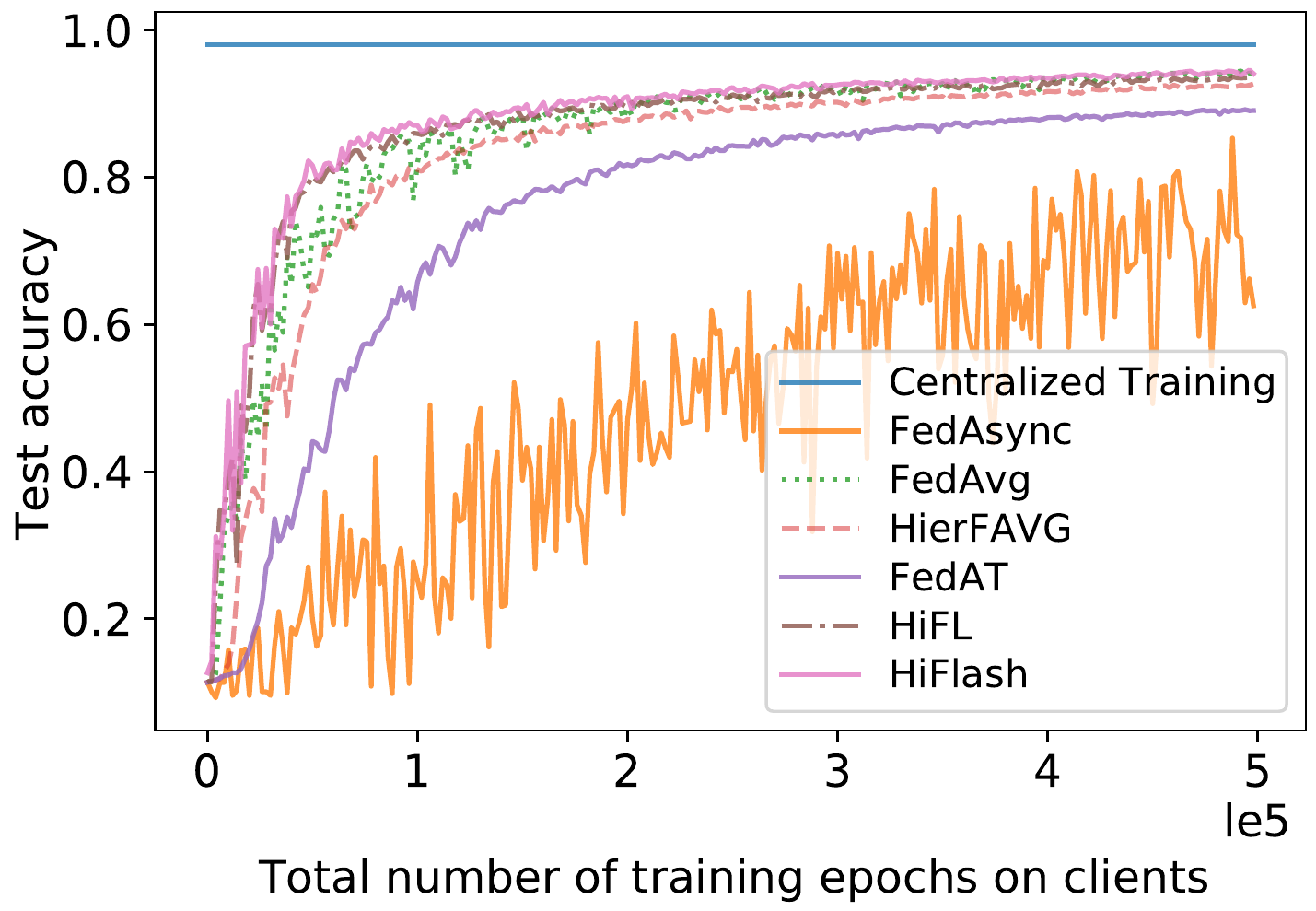}
	} 	 
	\subfigure[IID]{
		\includegraphics[width=0.315\linewidth]{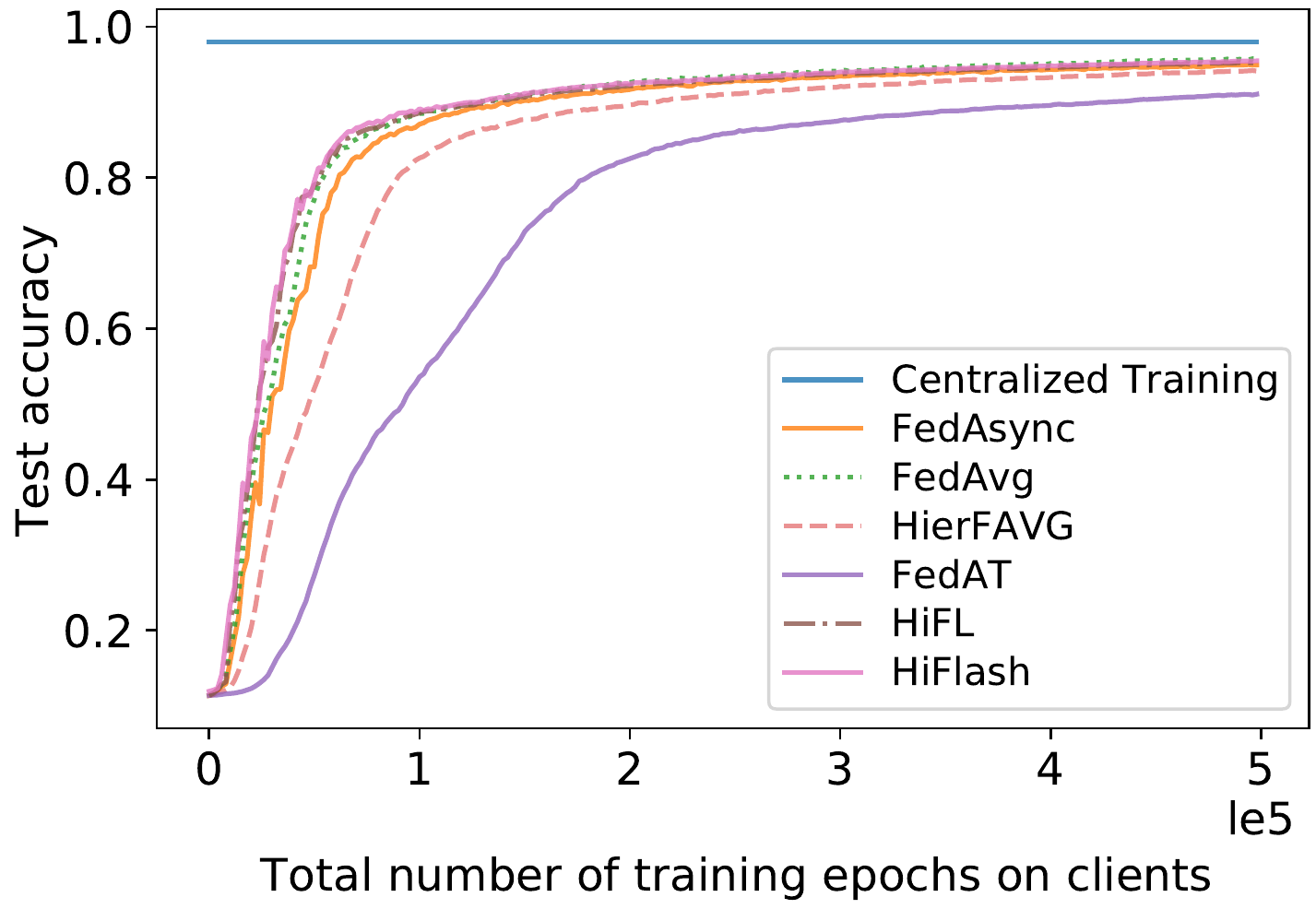}
	}  
	\vspace{-5pt}
	\caption{Test accuracy of MNIST dataset under different data distributions w.r.t the total number of training epochs on the clients.}
	\label{mnist_3dis}
	\vspace{-5pt}
\end{figure*}
\begin{figure*}[!t]
	\centering
	\subfigure[Non-IID(1)]{
		\includegraphics[width=0.315\linewidth]{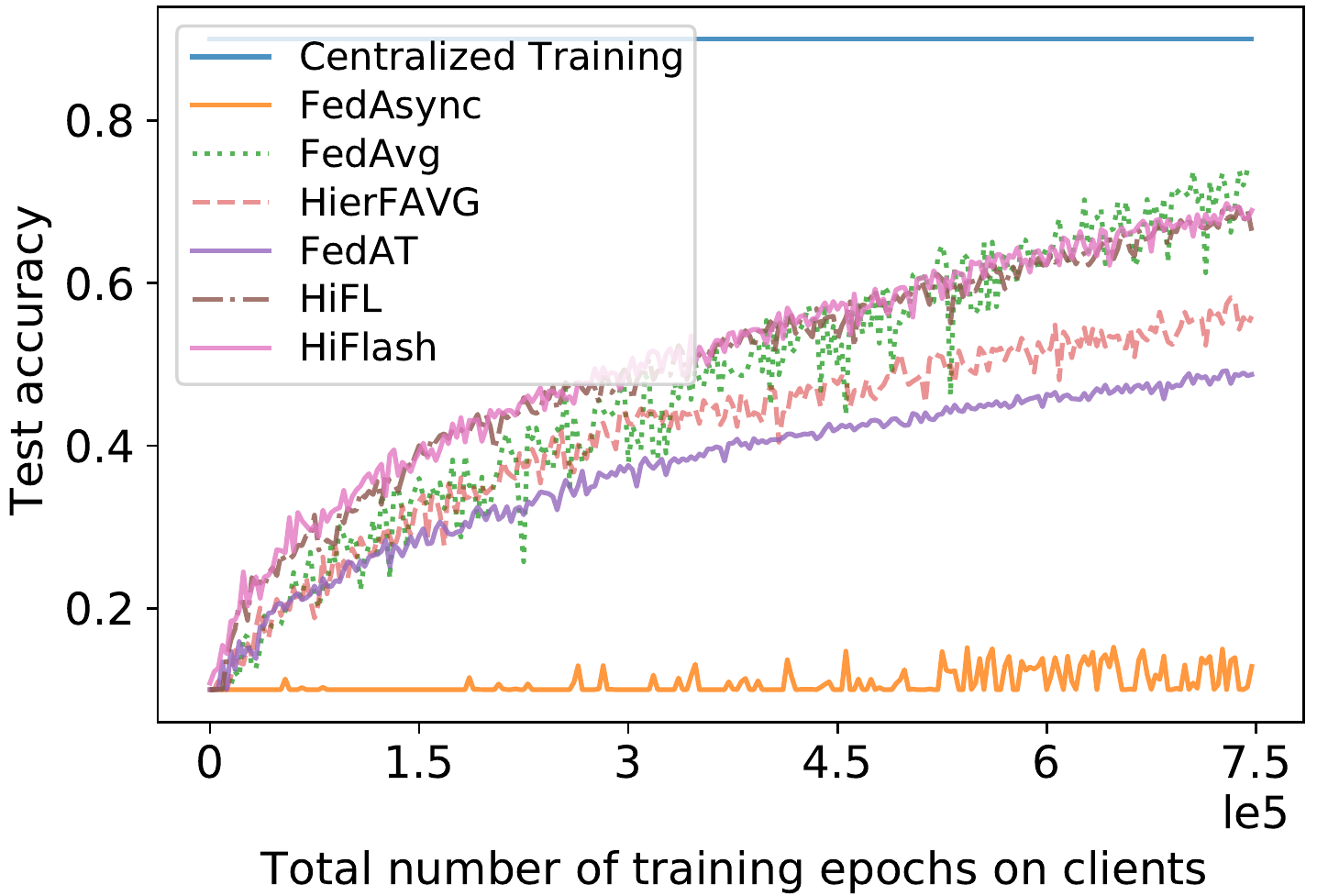}
	} 
	\subfigure[Non-IID(2)]{
		\includegraphics[width=0.315\linewidth]{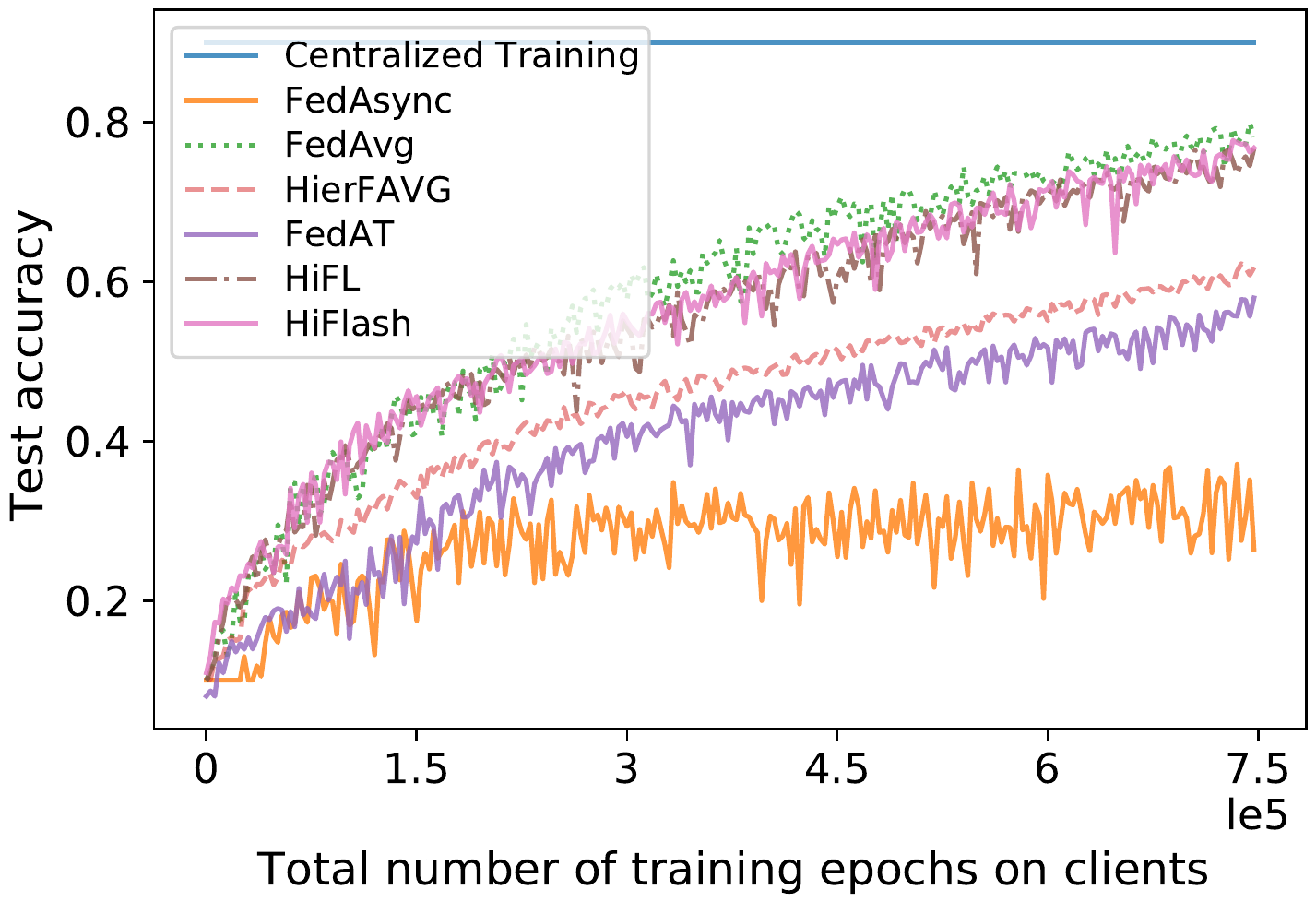}
	} 	 
	\subfigure[IID]{
		\includegraphics[width=0.315\linewidth]{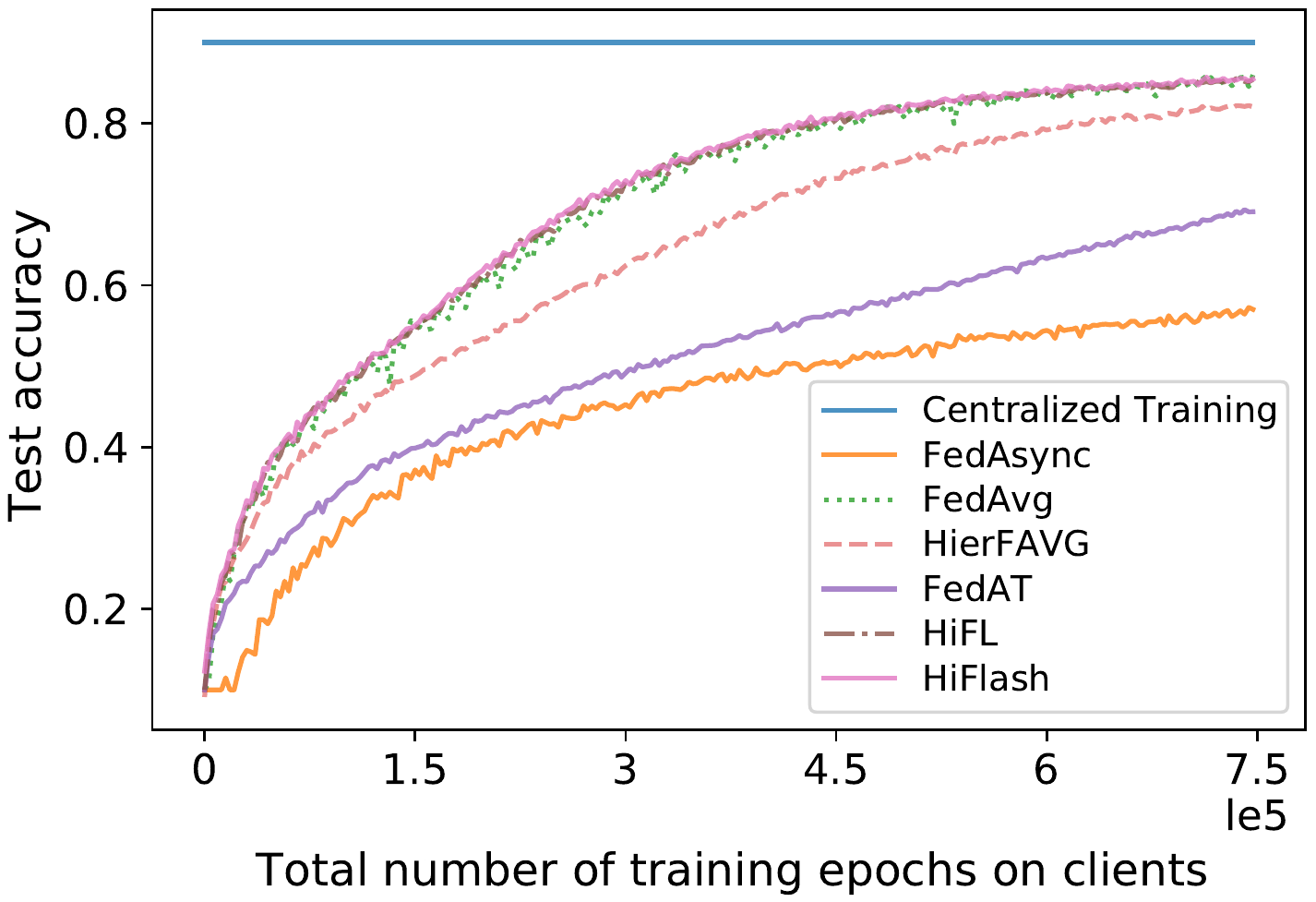}
	}  
	\vspace{-5pt}
	\caption{Test accuracy of CIFAR10 dataset under different data distributions w.r.t  the total number of training epochs on the clients.}
	\label{cifar10_3dis}
	\vspace{-10pt}
\end{figure*}

\begin{figure*}[!t]
	\centering
	\includegraphics[width=0.8\linewidth]{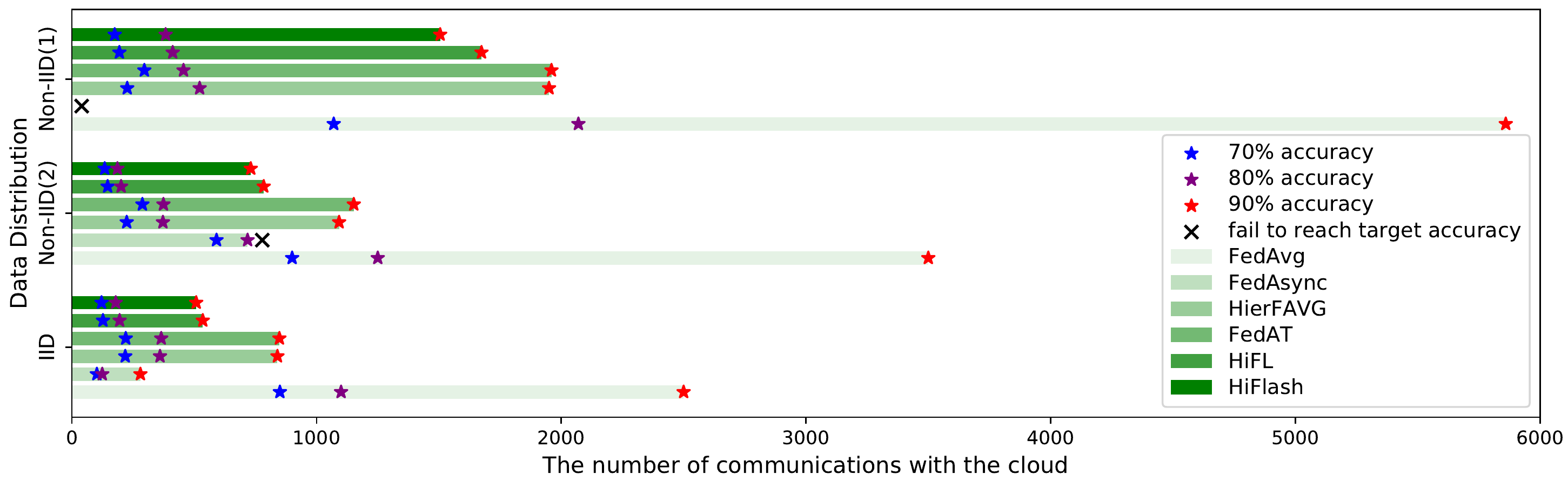}
	\vspace{-5pt}
	\caption{The number of needed communications between the edge nodes and cloud server to reach a target accuracy for HiFlash comparing with different FL methods under different levels of data heterogeneity of MNIST dataset.}
	\label{targetAcc_mnist}
	\vspace{-10pt}
\end{figure*}

\begin{figure*}[!t]
	\centering
	\includegraphics[width=0.8\linewidth]{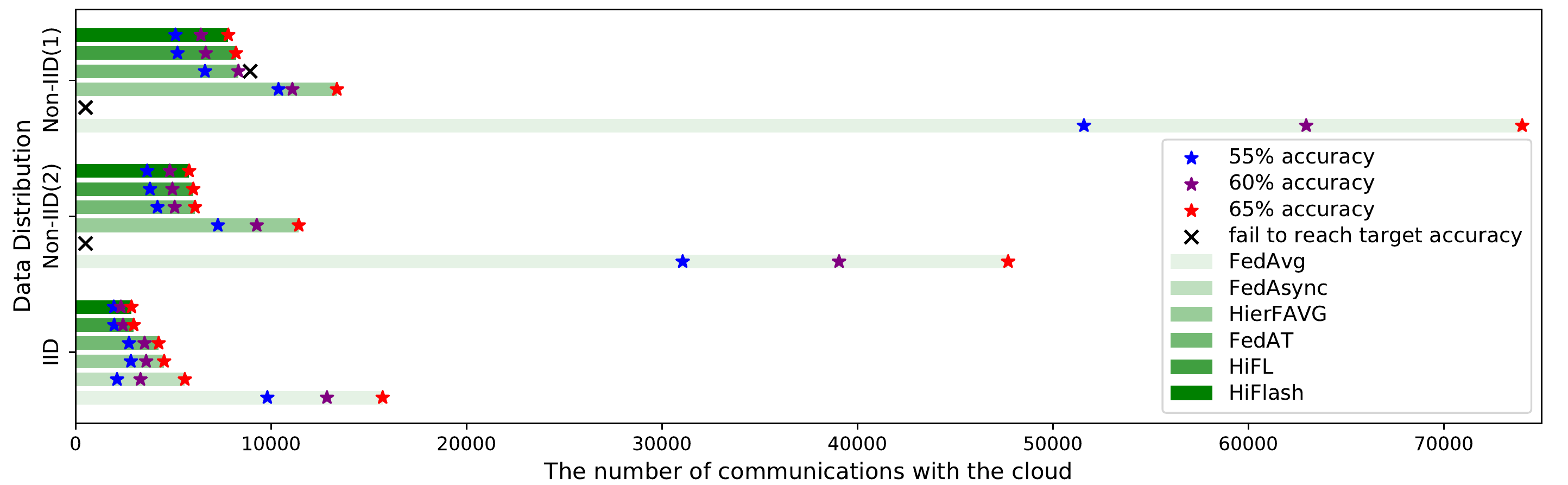}
	\vspace{-5pt}
	\caption{The number of needed communications between the edge nodes and cloud server to reach a target accuracy for HiFlash comparing with different FL methods under different levels of data heterogeneity of CIFAR10 dataset.}
	\label{targetAcc_cifar}
	\vspace{-10pt}
\end{figure*}

\subsection{Experimental Results}
\textbf{Performance evaluation for various hyperparameter settings.} 
As the global model updating of HiFL and HiFlash is controlled by $\alpha_{\tau}$ which is related to the initial weight of edge model $\alpha$ and penalty coefficient $\upsilon$, we first evaluate the test accuracy of HiFL under different settings of $\alpha$ and $\upsilon$. As shown in Fig. \ref{alpha_tau}, HiFL is robust and can converge within 1,000 communication rounds under the non-IID(2) data distribution with different mixing hyperparameter settings. However, a too large or too small value of $\alpha$ will result in a slow convergence speed. For example, when $\alpha$ is too large (e.g., $\alpha = 0.95$, $\alpha = 1.0$), the current global model fails to retain information about the global model from the previous round. While a smaller $\alpha$ (e.g., 0.5) will prevent the global model learning from the newly uploaded edge model. 

Hence, we adopt grid search to find a proper choice for $\alpha$ and $\upsilon$. We can see from Fig. \ref{alpha_tau} that the global model converges fast when $\alpha = 0.9$ and $\alpha = 0.7$, however, the test accuracy becomes more fluctuating with $\alpha = 0.9$ due to the deviation of the edge model. Besides, to deal with model staleness, a proper $\upsilon$ (e.g., 0.99) is effective for fast and stable model convergence. As the convergence results of different datasets under various mixing hyperparameter settings are similar, we set $\alpha = 0.7$ and $\upsilon = 0.99$ for the following experiments.

\textbf{Model accuracy and computation efficiency evaluation.} 
Considering the fact that hierarchical FL executes more local computations in one global round to reduce the costly communication with the cloud, we propose to evaluate the test accuracy with respect to the total number of training epochs on clients. As depicted in Fig. \ref{mnist_3dis} and Fig. \ref{cifar10_3dis}, we investigate the model accuracy and computation efficiency of different training methods under three kinds of data heterogeneity for MNIST and CIFAR10 datasets, respectively. As the centralized training method collects all the data to the cloud for model learning, it does not incur any computation cost on devices. Hence, we only use the test accuracy of centralized learning to provide an upper bound of model accuracy for other comparing methods.

We can see that by incorporating the merits of synchronous and asynchronous model aggregation and dampening the negative effect of model staleness, HiFL and HiFlash can achieve comparable training performance with FedAvg method in IID cases. For Non-IID cases, HiFL and HiFlash perform slightly less well than FedAvg when comparing test accuracy with respect to total number of training epochs on clients. This is because that the multiple rounds of client-edge aggregation in HiFL and HiFlash might lead to some degree of gradient divergence, and hence degrade the model performance. Moreover, as HiFL and HiFlash are designed with asynchronous model aggregation, they inevitably suffer from staleness effect, comparing with FedAvg.

HiFL and HiFlash perform better than other hierarchical FL methods (e.g., HierFAVG). Since hierarchical FL is designed to reduce the costly communication at the price of more local computations, HierFAVG, the extension of FedAvg in the hierarchical setting, is computationally inefficient comparing with FedAvg as shown in  Fig. \ref{mnist_3dis} and Fig. \ref{cifar10_3dis}. This is because that HierFAVG performs fewer edge-cloud aggregations than FedAvg for the same amount of local training epochs. While with the asynchronous update mechanism which well balances the global model and the uploaded edge model in HiFL and HiFlash, we can see that the performance gap between HiFL and FedAvg narrows significantly, comparing with that between HierFAVG and FedAvg, indicating that the asynchronous aggregation in HiFL and HiFlash is more computationally efficient than other hierarchical FL schemes.

HiFL and HiFlash perform better than other synchronous or asynchronous based methods. For example, asynchronous FL methods (e.g., FedAsync and FedAT) have lower test accuracy than HiFL and HiFlash, since they ignore the negative impacts of biased data distribution and model staleness on the cloud model accuracy. The sharp oscillation in the curves of FedAsync algorithm attributes to the following two reasons: (1) the global model in FedAsync algorithm is updated once one client uploads its updated model without waiting for stragglers. This kind of asynchronous aggregation induces much uncertainty into the performance of the resulting global model, especially in Non-IID cases. While other algorithms fuse different client models to ensure the generalization ability of the global model; and (2) the staleness effect makes the convergence of FedAsync slower and causes the performance instability when facing large staleness. 
\begin{figure*}[t!]
	\centering
	\begin{minipage}[t]{0.31\textwidth}
		\centering
		\includegraphics[width=1.0\textwidth]{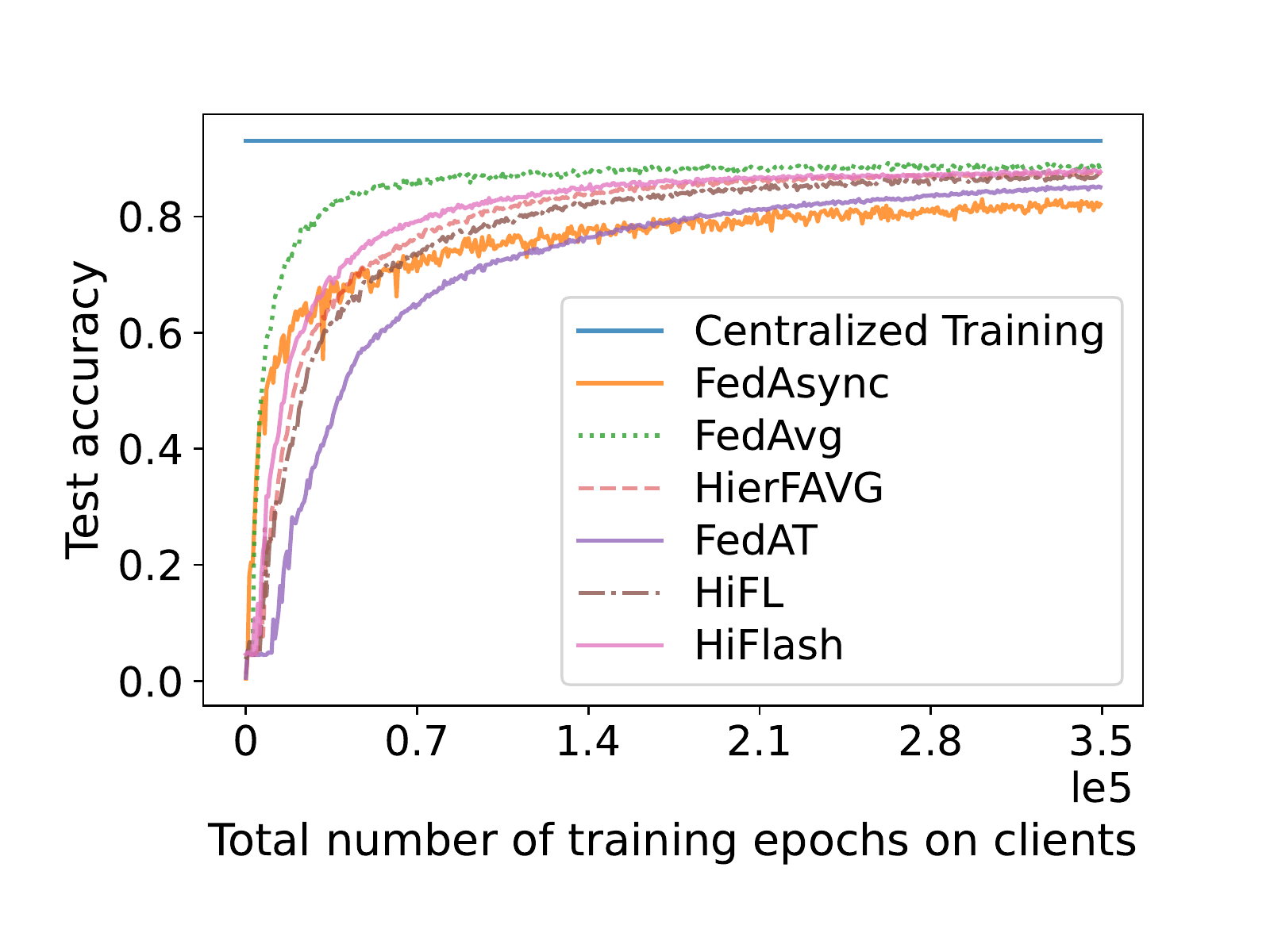}
		\vspace{-15pt}
		\caption{Test accuracy of FEMNIST dataset w.r.t. the total number of training epochs on the clients.}
		\vspace{-10pt}
		\label{acc_femnist}
	\end{minipage}
	\hspace{0in}
	\centering
	\begin{minipage}[t]{0.32\textwidth}
		\centering
		\includegraphics[width=1.0\textwidth]{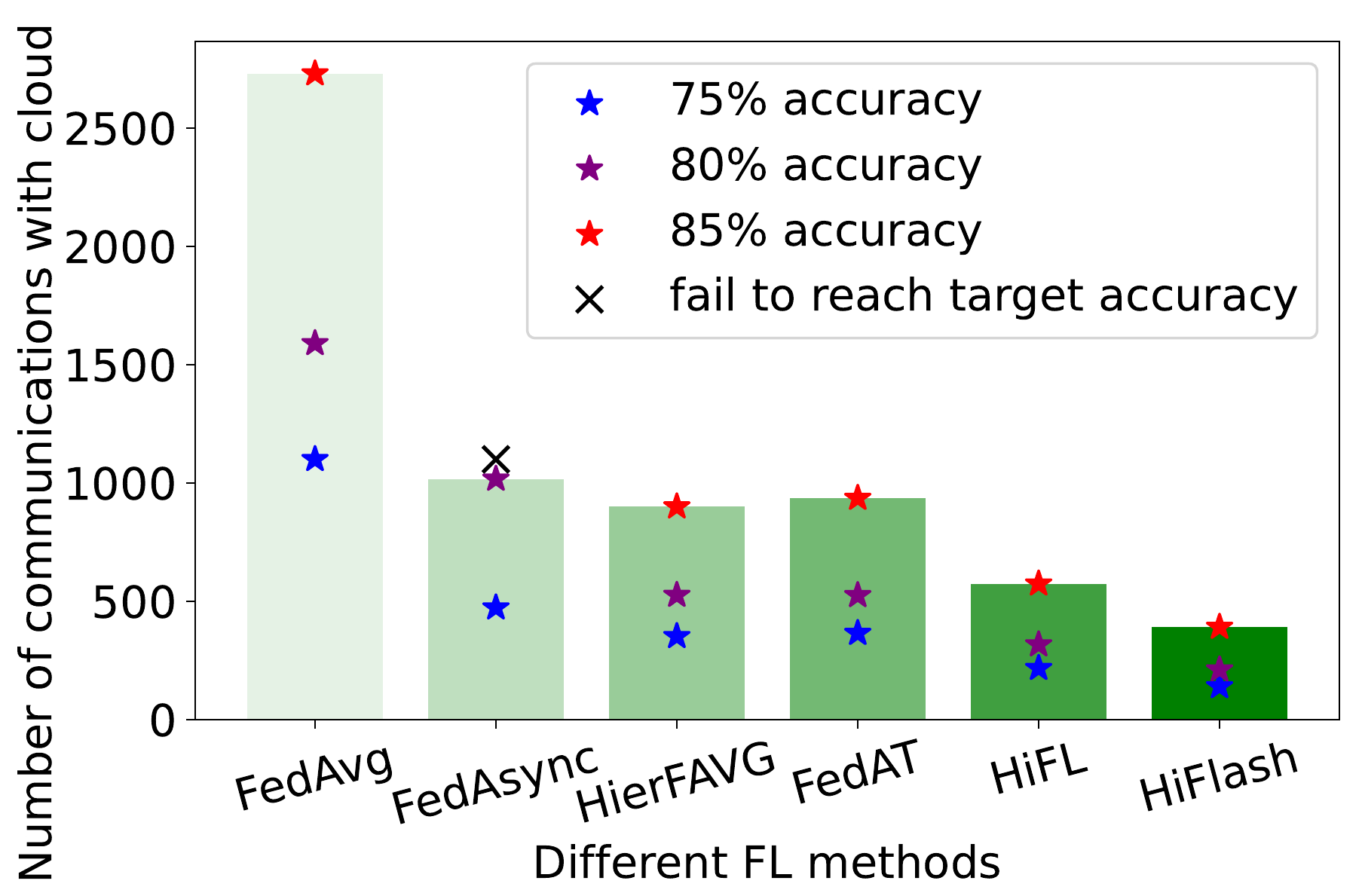}
		\vspace{-15pt}
		\caption{Number of needed communications between edge nodes and cloud server to reach a target accuracy under FEMNIST.}
		\vspace{-10pt}
		\label{comm_femnist}
	\end{minipage}
	\hspace{0in}
	\centering
	\begin{minipage}[t]{0.32\textwidth}
		\centering
		\includegraphics[width=1.0\textwidth]{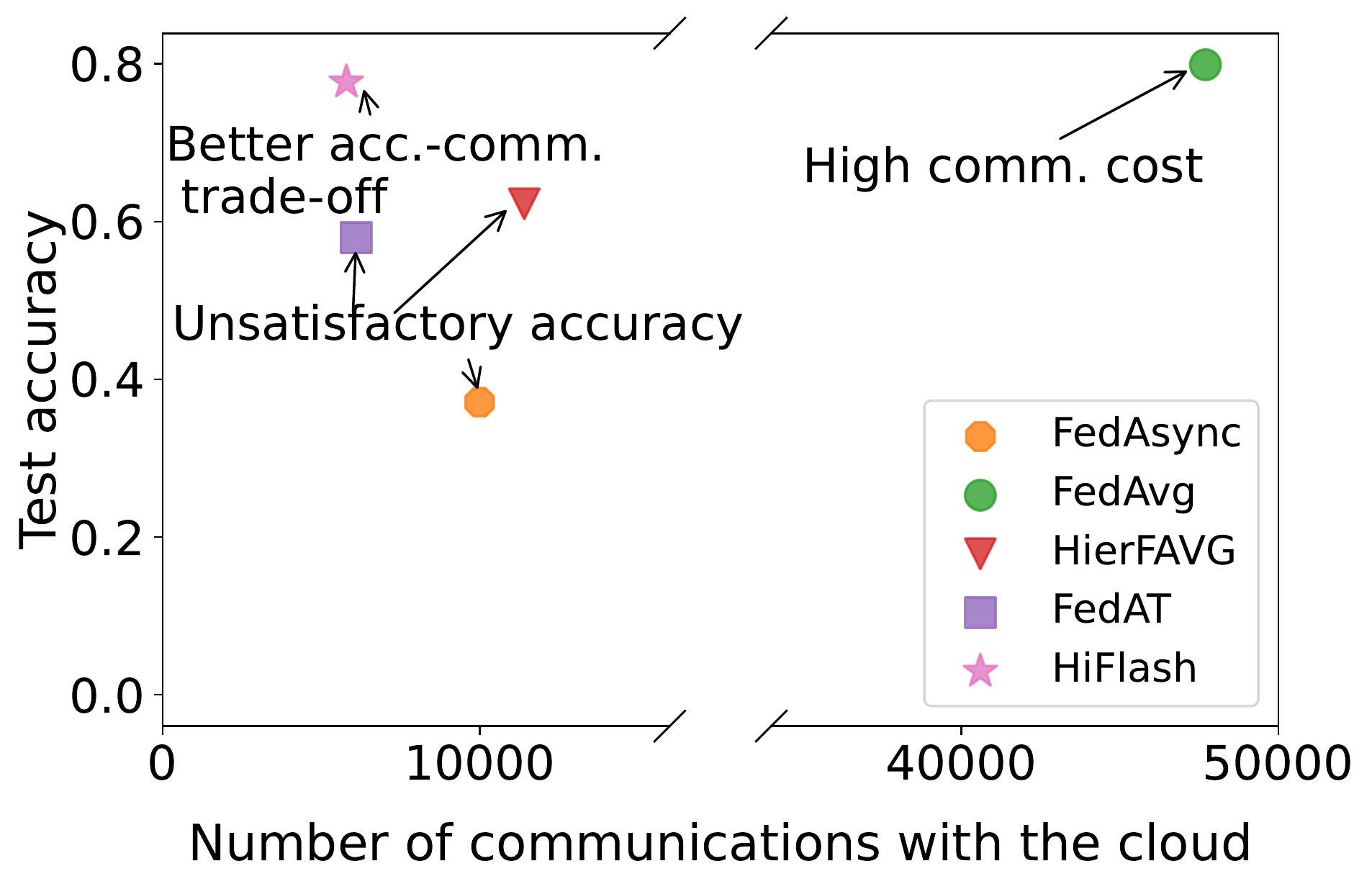}
		\vspace{-15pt}
		\caption{HiFlash strikes a nice accuracy-communication trade-off comparing with other FL based approaches.}
		\vspace{-10pt}
		\label{trade-off}
	\end{minipage}
	\hspace{0in}
	\vspace{-5pt}
\end{figure*}

\begin{figure*}[t!]
	\centering
	\begin{minipage}[t]{0.315\textwidth}
		\centering
		\includegraphics[width=1.0\textwidth]{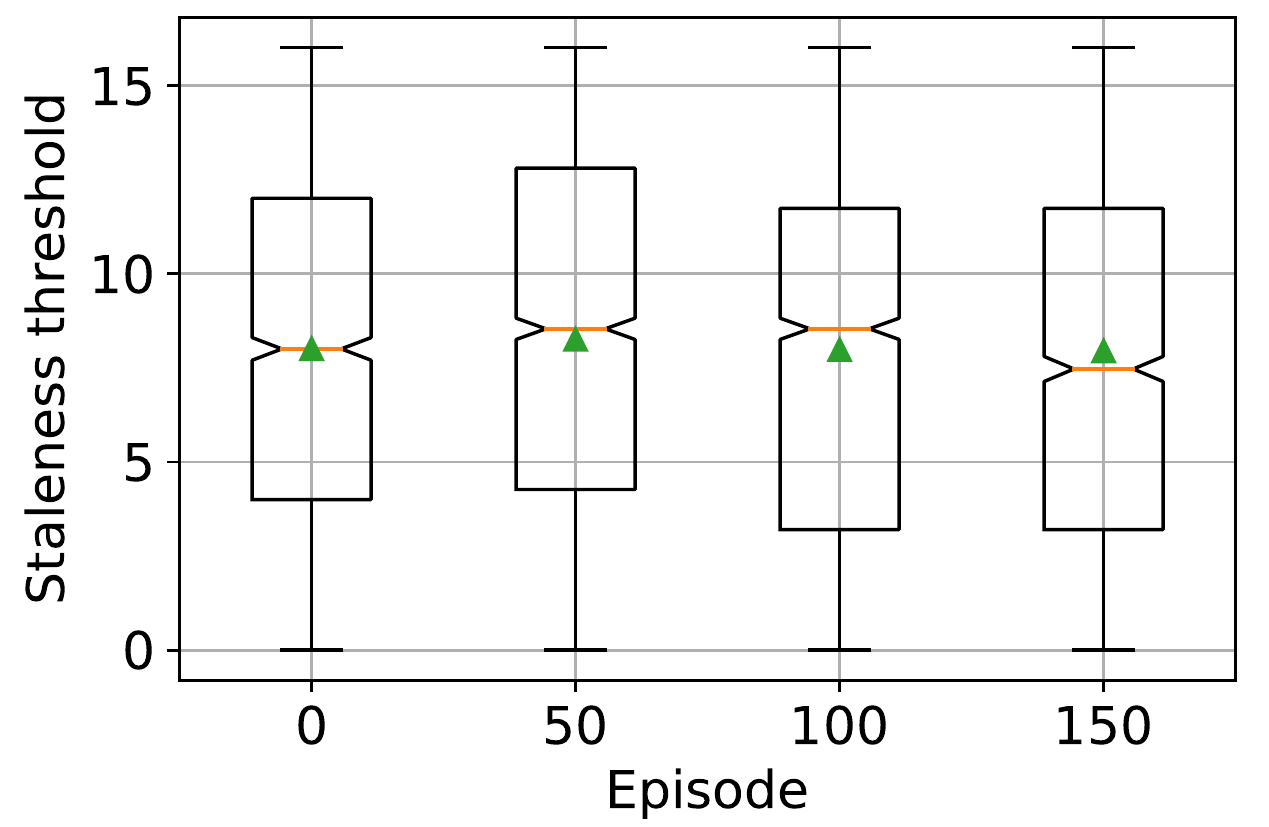}
		\vspace{-18pt}
		\caption{The policy improvement process for adaptive staleness control with the increasing of DRL training episodes.}
		\vspace{-10pt}
		\label{episode_threshold}
	\end{minipage}
	\hspace{0in}
	\centering
	\begin{minipage}[t]{0.315\textwidth}
		\centering
		\includegraphics[width=1.0\textwidth]{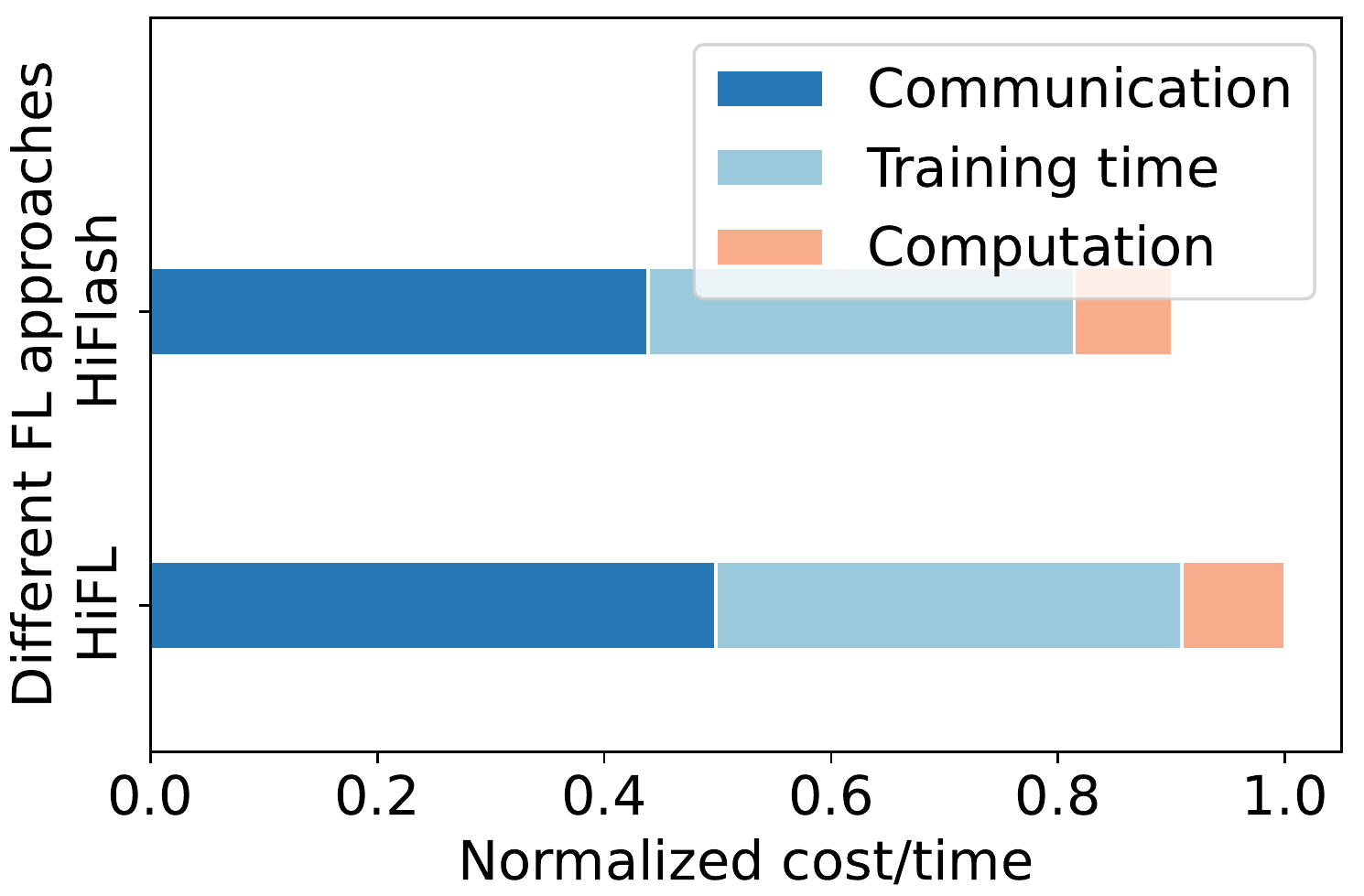}
		\vspace{-18pt}
		\caption{The total system costs for HiFL and HiFlash.}
		\vspace{-10pt}
		\label{normalizedCostCompare}
	\end{minipage}
	\hspace{0in}
	\centering
	\begin{minipage}[t]{0.315\textwidth}
		\centering
		\includegraphics[width=1.0\textwidth]{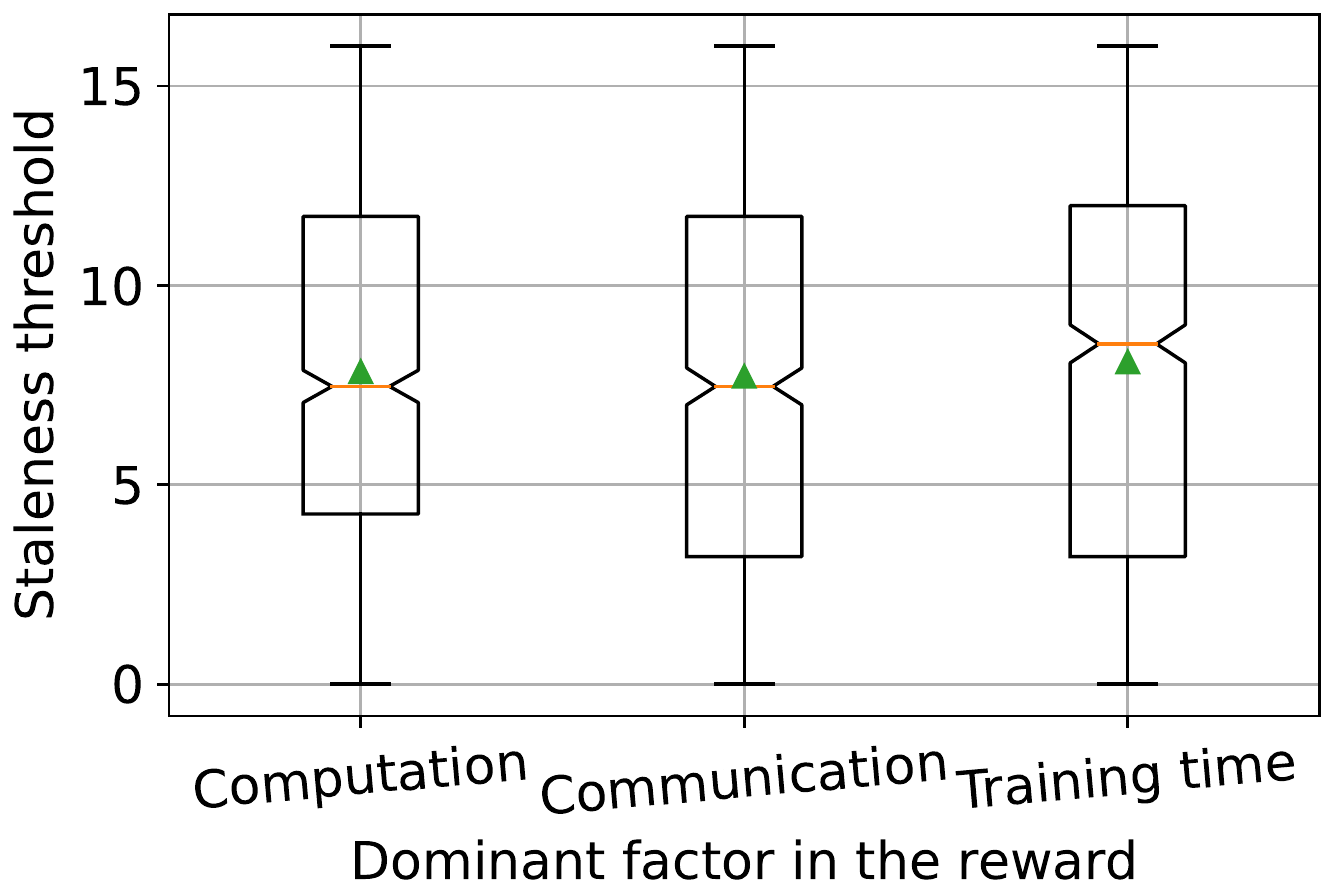}
		\vspace{-18pt}
		\caption{Policy differences when facing various reward designs.}
		\vspace{-10pt}
		\label{dominant_factor}
	\end{minipage}
	\hspace{0in}
	\vspace{-5pt}
\end{figure*}

\textbf{Communication efficiency evaluation.} We define the number of communications in FL process as the total communication number between edge nodes (or clients in two-layer FL frameworks) and the cloud server for model exchange. A smaller number of communications with the cloud indicates a smaller data size of models transferred to the cloud. To evaluate communication efficiency of HiFlash, we investigate the number of communications between the edges and the cloud to reach a target accuracy for all the FL based approaches. 

As shown in Fig. \ref{targetAcc_mnist}, for MNIST dataset, the required communication numbers for different methods grow with the increase of target accuracy and data heterogeneity of clients. Except FedAsync algorithm, our proposed HiFL scheme is the most communication-efficient than other FL based methods regardless of data distribution and target accuracy. For example, HiFL can reduce the communication numbers by up to $31.9\%$ than FedAT, $28.2\%$ than HierFAVG and $77.6\%$ than FedAvg on MNIST dataset with Non-IID(2) distribution and target accuracy of $90\%$. Although FedAsync can reach a target accuracy with fewer communication numbers than our proposed HiFL method in IID setting for MNIST dataset, it fails to deal with the data heterogeneity (e.g., Non-IID(2) and Non-IID(1) cases) inherent in the participating clients. For example, FedAsync can not achieve the target accuracy of $90\%$ under Non-IID(2) distribution and even fails to reach the target accuracy of $70\%$ in Non-IID(1) case, indicating that FedAsync is not applicable in realistic FL scenarios where data is distributed in a Non-IID fashion. 

The hierarchical FL methods (e.g., HiFL, HierFAVG) significantly reduce the costly communications with the cloud due to the client-edge aggregations. Moreover, the enhanced HiFlash framework, equipped with adaptive staleness control, can further accelerate the model training process and reduce the communication rounds with the cloud comparing with HiFL (e.g., $10\%$ communication round reduction under Non-IID(1) data distribution case). This result is consistant with the convergence analysis that the model will converge within fewer communication rounds $T_c$ by controlling $\tau$ in a smaller value. Thus, the client-edge aggregation and staleness control contribute to the communication efficiency of HiFL and HiFlash.

As for the comparison of FedAvg and HierFAVG, there are 10 clients communicating with the cloud in each training round for FedAvg, while 5 edges communicate with the cloud in HierFAVG method. Moreover, HierFAVG uses more local computation on the clients in each round to decrease the number of global training rounds, thus, HierFAVG is much better than FedAvg in terms of communication cost (e.g., the number of communications with the cloud).

For a more complicated dataset (i.e., CIFAR10), HiFlash can reach different target accuracies with the smallest communication rounds in all data distribution situations, comparing with all the FL based methods. As depicted in Fig. \ref{targetAcc_cifar}, HiFlash requires $6403$ communication numbers with the cloud to reach the target accuracy of $60\%$ in Non-IID(1) data distribution scenario, which is $23.07\%$,  $42.23\%$ and $89.82\%$ smaller than FedAT, HierFAVG, and FedAvg, respectively. It shows that HiFlash outperforms current existing hierarchical FL algorithms, no matter asynchronous based FedAT or synchronous based HierFAVG method.

\textbf{Evaluation results on FEMNIST dataset.} We also evaluate the performance of HiFlash under FEMNIST dataset where the data distributions of clients are naturally non-IID (in feature distribution, label distribution and quantity distribution). As shown in Fig. \ref{acc_femnist}, HiFlash still performs better than other hierarchical FL methods (e.g., HierFAVG), which is similar with the experimental results under MNIST dataset. Moreover, the number of communications with the cloud of HiFlash approach is the smallest comparing with other FL methods, which can be seen in Fig. \ref{comm_femnist}. The evaluation results show that our HiFlash approach can be applied in real-world datasets with skews in both label distribution and feature distribution.

\textbf{Accuracy-communication trade-off.} To clearly show the superiority of HiFlash, we further investigate the accuracy-communication trade-off for different FL methods. Giving CIFAR10 dataset with Non-IID(2) distributions as an example, we plot the highest model accuracy and the communication rounds with the cloud in Fig. \ref{trade-off}. We can see that the popular FedAvg approach suffers from high communication cost, while asynchronous based FedAsync approach and hierarchical FL methods (e.g., HierFAVG and FedAT) fails to achieve a satisfactory accuracy. In contrary, HiFlash is able to strike a nice balance between model accuracy and communication efficiency. Specifically, HiFlash significantly reduces the communication cost (e.g., $87\%$ than FedAvg) with a slight model accuracy degradation (e.g., $2.1\%$). When comparing with hierarchical FL approaches (e.g., FedAT and HierFAVG), HiFlash is able to achieve more than $5\%$ communication cost reduction and $16\%$ model accuracy improvement. Furthermore, the superiority of HiFlash can be amplified as data heterogeneity increases (see Fig. \ref{targetAcc_mnist} and Fig. \ref{targetAcc_cifar}). 

\textbf{The effect of staleness threshold control.} The fast model convergence speed and high communication efficiency achieved by HiFlash are attributed to the well design of DRL-based adaptive staleness control that makes a wise decision according to the past experiences and current environment. As illustrated in Fig. \ref{episode_threshold}, the DRL agent for adaptive staleness control improves the  staleness threshold decision policy unremittingly as it interacts with the FL environment and learns from the DRL training episodes. We adopt boxplot to graphically depict the five-number summary of the distribution of the staleness threshold decisions in different training episodes, which consists of the smallest observation, lower quartile, median, upper quartile and largest observation. The upper quartile, median and lower quartile make up a box with compartments. The spacings between different parts indicate the variance and skew in the data distribution and the mean of staleness thresholds in shown with the green triangles. The decision policy in episode 0 is the random staleness control policy adopted in HiFL while the improved policy in episode 150 is utilized by the HiFlash approach. We can learn from the skewed data distribution that more decisions made in HiFlash choose a smaller staleness threshold compared with random decision policy, resulting in communication-efficient model training. 

To compare the system costs (including computation cost of the clients, communication cost of the edges and training time of the cloud) of HiFL and HiFlash, we normalize the system cost to $[0,1]$ by simply dividing the biggest value.  With this normalization method, the computation cost and communication cost are scaled down accordingly. As shown in Fig. \ref{normalizedCostCompare}, the normalized cost of HiFL is 1, indicating that HiFL brings high system cost. While the lower system cost of HiFlash is credited to the effectiveness of our adaptive staleness control strategy design. It is worthnoting that although the DQN training in HiFlash brings additional computation overhead, it can be conducted on the cloud server with sufficient computation resources in an offline manner.

\begin{figure*}[t!]
	\centering
	\begin{minipage}[t]{0.32\textwidth}
		\centering
		\includegraphics[width=1.0\textwidth]{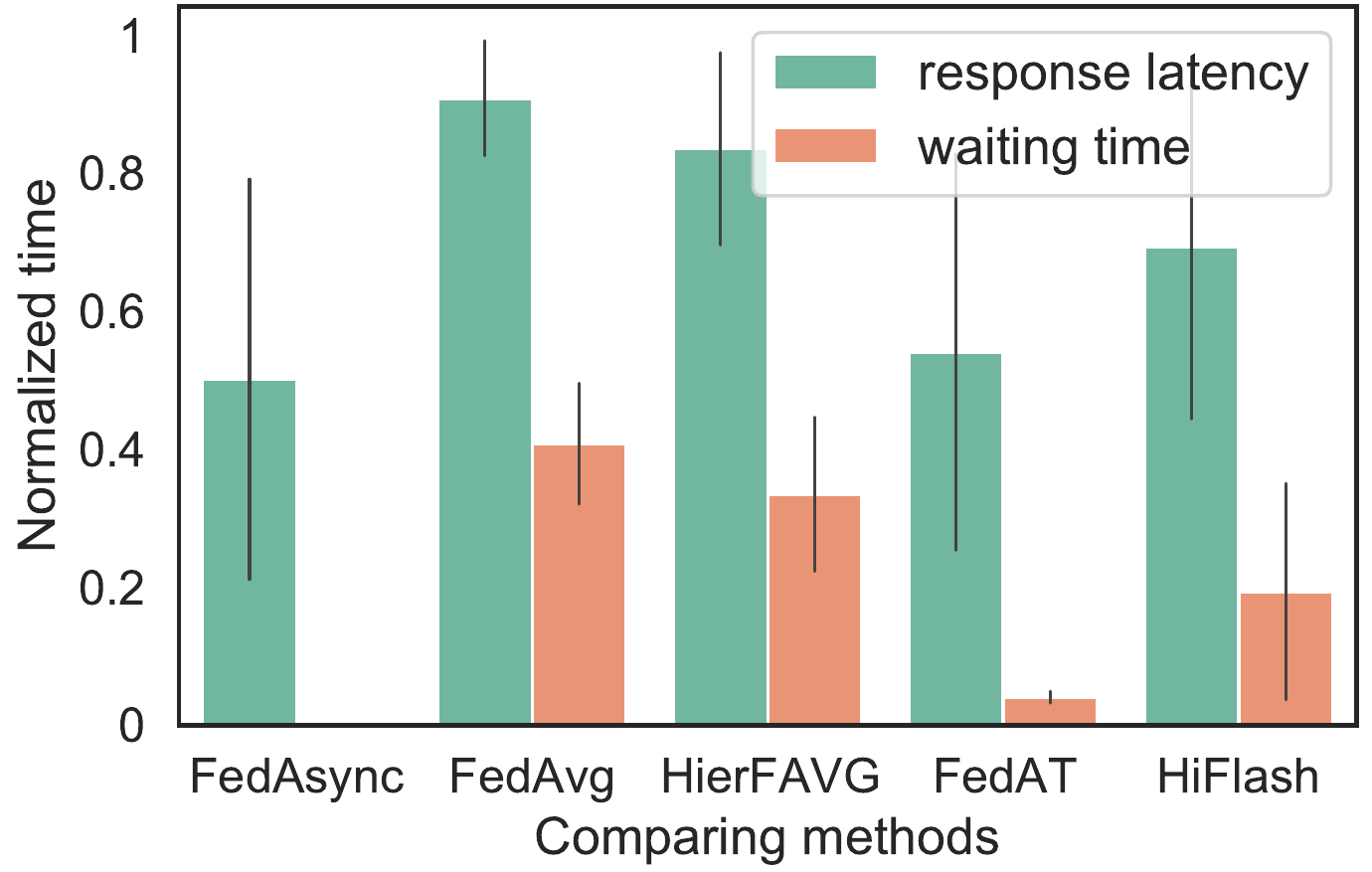}
		\vspace{-17pt}
		\caption{Average response latency and waiting time for different FL methods.}
		\vspace{-10pt}
		\label{latency}
	\end{minipage}
	\hspace{0in}
	\centering
	\begin{minipage}[t]{0.315\textwidth}
		\centering
		\includegraphics[width=1.0\textwidth]{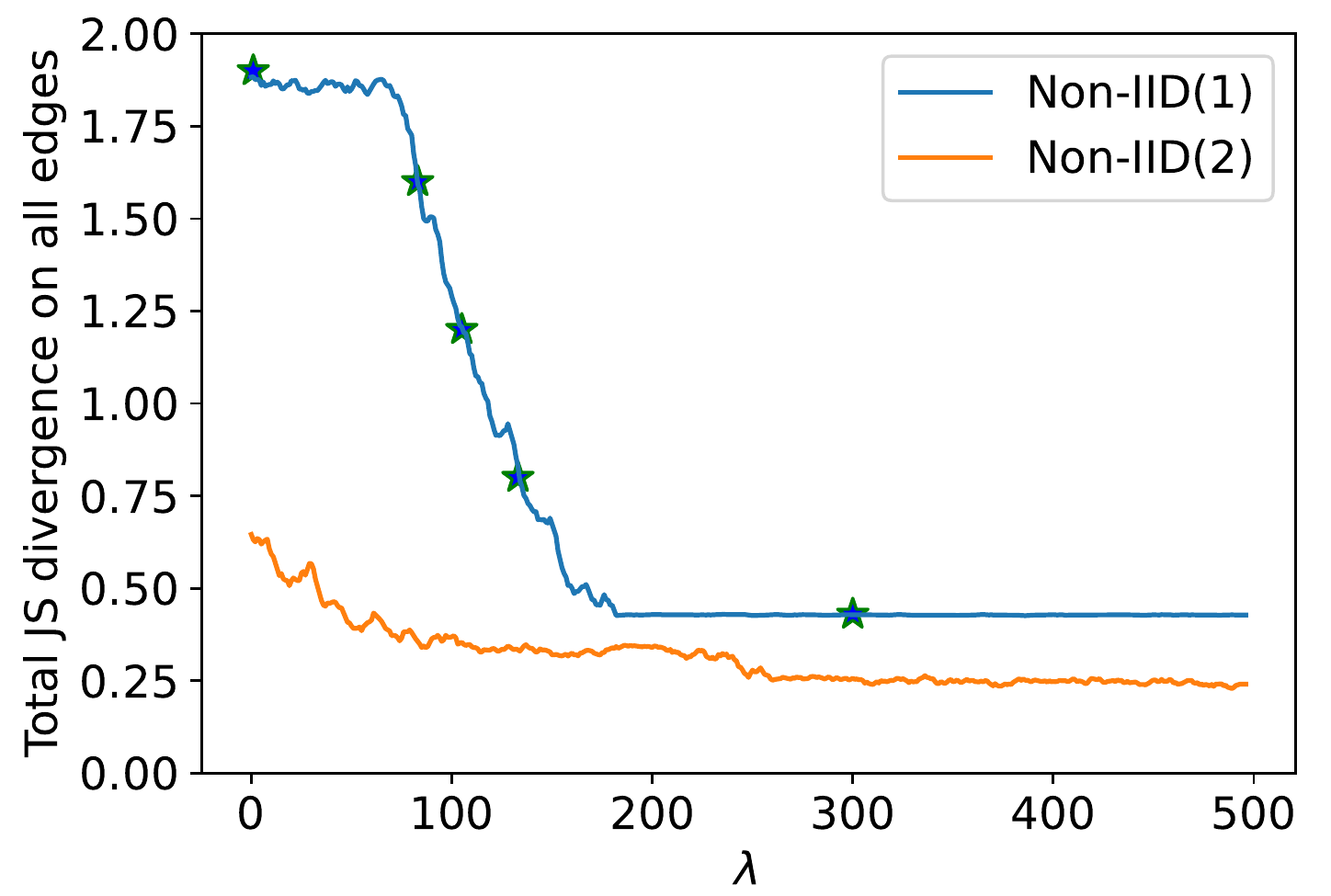}
		\vspace{-17pt}
		\caption{JS divergence with varying $\lambda$ under different data heterogeneity.}
		\vspace{-10pt}
		\label{varyingMu1}
	\end{minipage}
	\hspace{0in}
		\begin{minipage}[t]{0.32\textwidth}
		\centering
		\includegraphics[width=1.0\textwidth]{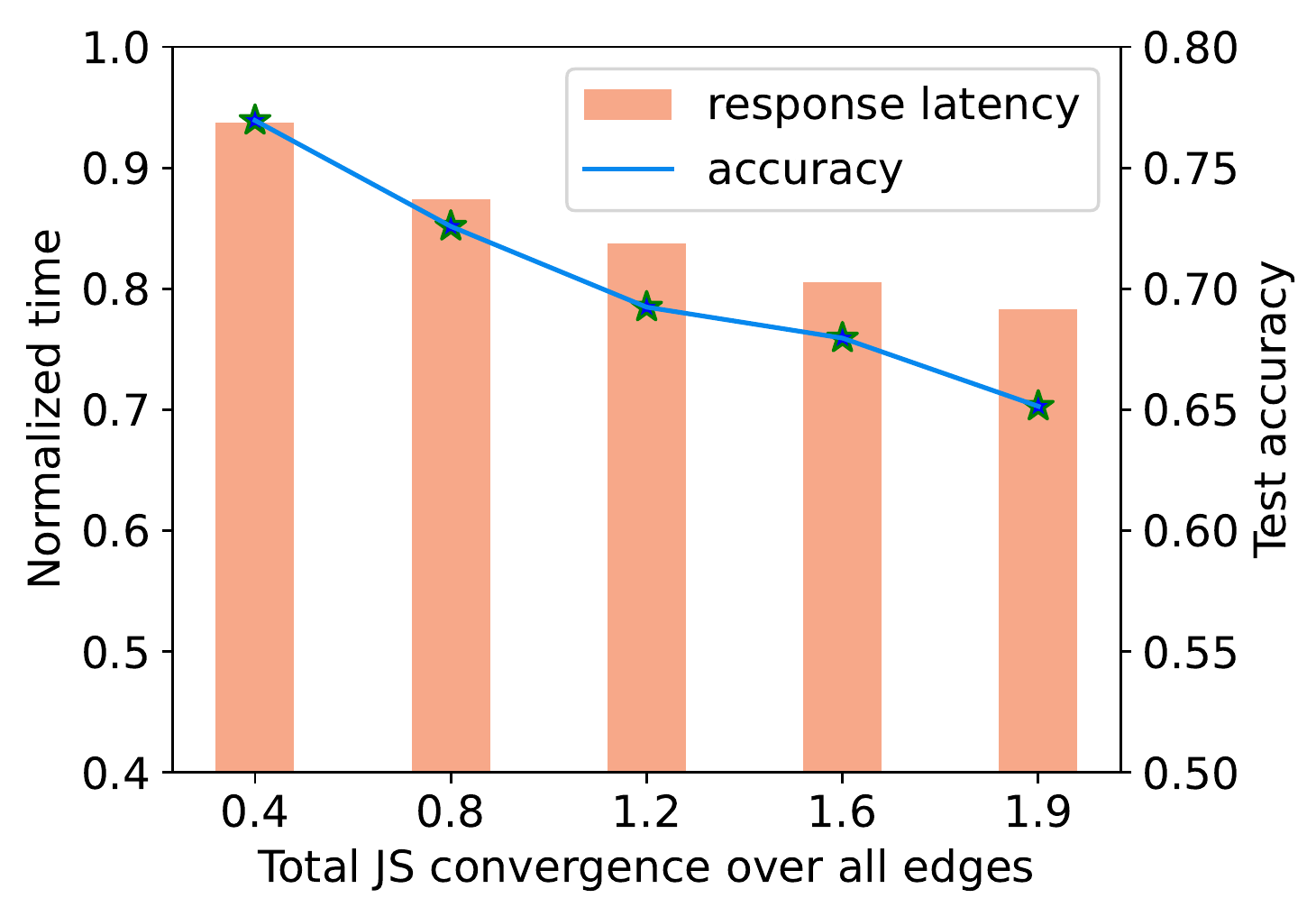}
		\vspace{-17pt}
		\caption{Policy differences when facing various reward designs.}
		\vspace{-10pt}
		\label{varyingMu2}
	\end{minipage}
	\hspace{0in}

\end{figure*}
We also examine the policy differences with various reward designs by assigning different weights to computation cost, communication cost and training time (adjusting the values of $\sigma_{1}$ and $\sigma_{2}$). As shown in Fig. \ref{dominant_factor}, when assigning higher weights to resource cost (e.g., computation and communication cost), the DRL agent for staleness control tends to choose smaller threshold in fear of severe staleness effect. While for a higher weight of training time, more decisions with bigger staleness threshold are made to facilitate model parallelism.

\textbf{The effect of heterogeneity-aware client-edge association.} Besides, we also evaluate the response latency $L^{m,k}$ and waiting time $L^m_{waiting}$ of the edge-associated clients in the client-edge aggregation phase. For HiFlash, we set $\lambda = 300$ (will be discussed in next paragraph) to form a more label-balanced dataset for each edge node. Thus, the average response latency may be longer due to the trade-off between data heterogeneity and latency reduction. As shown in Fig. \ref{latency}, the average response latency and waiting time of FedAsync are the lowest since there is no need to wait for stragglers. FedAT, an asynchronous hierarchical FL scheme similar with our proposed HiFL, also has lower response latency and waiting time. However, it only focuses on latency reduction in client-edge association procedure without the consideration of data heterogeneity mitigation. Thus, both FedAsync and FedAT result in degraded accuracy as in Fig. \ref{mnist_3dis}-\ref{cifar10_3dis} and  more communications as in Fig. \ref{targetAcc_mnist}-\ref{targetAcc_cifar}. While our proposed HiFlash can enforce a trade-off between the response latency and data heterogeneity, achieving a satisfactory model performance. 

We further investigate the total JS divergence on all the edges nodes, the resulted average response latency, and the model accuracy of HiFlash with varying $\lambda$. As depicted in Fig. \ref{varyingMu1}, a bigger $\lambda$ denotes that we are more concerned about the data distributions on the edge nodes compared with response latency, thus the total JS divergence decreases with the increasing $\lambda$ in both Non-IID(1) and Non-IID(2) cases. In Fig. \ref{varyingMu2}, a biased edge data distribution (larger JS divergence) will cause significant accuracy degradation. For example, the achievable accuracy drops from $77\%$ to $65\%$ when the total JS divergence over all edge nodes increases from $0.4$ to $1.9$. By controlling the parameter $\lambda$, our proposed HiFlash can be flexible for edge nodes to strike a balance between response latency and model accuracy.

\section{Related Work}
\label{SectionRelatedWork}
\subsection{Communication-Efficient Federated Learning}
Classical two-layer FL frameworks (e.g., FedAvg \cite{mcmahan2017communication} and FedAsync \cite{Xie2019Asyn}) inevitably suffer from excessive communication overhead and network congestion in large-scale distributed machine learning, due to massive model exchanges between clients and central server. To reduce the bits on gradient exchanges in FL, techniques such as neural network pruning \cite{jiang2019model}, weight quantization \cite{alistarh2017qsgd}, message sparsification \cite{luping2019cmfl} and knowledge distillation \cite{ahn2019wireless} focus on ML model compression to reduce the amount of transmitted information while maintaining the high learning performance. 

To further improve communication efficiency, hierarchical FL is proposed by introducing an edge layer, which leverages edge nodes as intermediaries to perform partial model aggregation with efficient client-edge communication, and thus relieves core network transmission overhead in the cloud server\cite{luo2020hfel}. For example, Liu et al. propose HierFAVG that performs two level of synchronous model aggregation by extending the conventional FedAvg algorithm to the hierarchical setting \cite{liu2020client}. However, a severe straggler problem would incur in HierFAVG due to the nature of synchronous model aggregation. Chai et al. present FedAT, a novel FL method that synergistically combines synchronous intra-tier training and asynchronous cross-tier training to improve the convergence speed and reduce communication cost \cite{chai2020fedat}. Nevertheless, FedAT uses a weighted sum of all the latest edge models for global model update, which is different from our asynchronous aggregation mechanism. Moreover, it ignores the staleness effect, which is inevitable in asynchronous aggregation. 

\subsection{Model Staleness Control}
Staleness effect is a common challenge in the asynchronous model aggregation. Most of existing FL solutions tolerate staleness by dampening the impacts of stale result that is computed on an outdated global model version. For example, Zhang et al. propose a staleness-aware async-SGD algorithm in which the learning rate is modulated according to the gradient staleness \cite{zhang2016staleness}. Xie et al. use a weighted average for global model update and introduce a mixed hyperparameter to adaptively control the error caused by staleness \cite{Xie2019Asyn}. An gradient correction term is designed to compensate the staleness in \cite{zhu2021delayed}. Nevertheless, these approaches only focus on the negative impact of staleness on model accuracy and convergence speed. The system efficiency, such as computation/communication cost, and training time, is less considered.

A stale model may marginally contribute to the global model, but cause large resource waste and time consumption without timely terminating model training and uploading. 
To address this issue, FedSA \cite{chen2021fedsa}, a staleness-aware asynchronous FL algorithm sets a staleness threshold for each participating client based on its computing speed. However, this approach ignores the communication cost and the training time in the whole process. Besides, the staleness threshold in FedSA is fixed for each client, which can not well adapt to the realistic dynamic environment. Zhang et al. propose a clustered semi-asynchronous federated learning (CSAFL) approach \cite{9533794}, which alleviates the model staleness problem by dividing clients with different learning objectives into multiple groups and limiting the model delay. However, CSAFL aims to learn personalized models, i.e., different group models for different client groups, while the HiFlash approach has a different goal and aims to learn a common global model that merits the commonality of the global knowledge sharing based on all the local data generated on clients.

%Staleness control in asynchronous FL is similar with the client selection in synchronous FL, as they are both decision making problems. Hence, the multi-arm bandit based selection methods, deep reinforcement learning algorithms used in client selection problem can also be adopted in staleness control. However, client selection approaches in synchronous FL usually strike a balance among model accuracy, system efficiency and selection fairness. As staleness control problem does not need to ensure fairness for each threshold choice due to asynchronous aggregation which only involves one client, the client selection approaches (e.g., \cite{Stochastic_huang}, \cite{Towards_wang}) with fairness consideration can not be utilized in staleness control problem. Moreover, in client selection approaches, the server often needs to send the global model to all the clients for local loss estimation before making selection decisions (e.g., \cite{Towards_wang}, \cite{lai2021oort}), which is not required for staleness control.

\subsection{Client-Edge Association in Hierarchical FL}
The client-edge association strategy in hierarchical FL has a significant impact on model learning performance due to the data and resource heterogeneity across the dispersed clients \cite{chai2020tifl}. HierFAVG ignores the varying training speed of the distributed clients and randomly groups them into different clusters, which may prolong the communication time of each training round \cite{liu2020client}. Considering the straggler effect, both TiFL \cite{chai2020tifl} and FedAT \cite{chai2020fedat} divide the clients into different tiers based on the measured latency so that the resource heterogeneity can be mitigated \cite{chai2020tifl}. To tackle the data heterogeneity in FL, Duan et al. propose FedGroup, which clusters clients into multiple groups based on the cosine similarity of their parameter updates \cite{duan2020fedgroup}. In FedCluster, the authors provide several representative clustering approaches, including random uniform clustering, timezone-based clustering and availability-based clustering, to support for various application scenarios \cite{chen2020fedcluster}. Unfortunately, current client-edge association schemes only focus on one dimension of heterogeneity, without considering the trade-off between data and resource heterogeneities.

Existing convergence analysis of hierarchical FL schemes mainly focus on FL with two levels of synchronous model aggregations. For example, the convergence of HierFAVG measures two-level of non-IIDness (i.e., the client level and the edge level) for data distribution in the hierarchical system, and provides qualitative guidelines on picking the aggregation frequencies at two levels \cite{liu2020client}. While the theoretical analysis in \cite{lee2020accurate} further proves that reducing the non-IIDness at the edge level is important for the model convergence. Although FedAT \cite{chai2020fedat} is designed with synchronous client-edge aggregation and asynchronous edge-cloud aggregation, the asynchronous aggregation mechanism in FedAT is a weighted sum of all the latest edge models, which is different from our asynchronous update mechanism. Moreover, the convergence analysis of FedAT ignores the staleness effect and hence fails to provide insights for staleness effect alleviation. Nevertheless, our convergence analysis for HiFL considers both staleness introduced by asynchronous model aggregation and non-IIDness inherent in FL, and further draws some insights for staleness control and non-IIDness reduction to achieve a fast convergence rate and a low convergence bound.

\section{Conclusion}
\label{SectionConclusion}
In this paper, we resort to HiFL, a hierarchical FL approach that synergistically employs synchronous client-edge model aggregation and asynchronous edge-cloud model aggregation for communication-efficient model learning. Based on the convergence analysis of HiFL, we identify the controllable factors for model convergence and further advocate HiFlash, an enhanced HiFL with adaptive staleness control and heterogeneity-aware client-edge association, for large-scale deployment in reality. We propose a DRL-based staleness threshold decision algorithm for accurate and cost-efficient FL model learning. To tackle the inherent resource and data heterogeneity among clients, we design a heterogeneity-aware client-edge association strategy that strikes a nice balance between communication latency and the heterogeneity of edge data distributions. Our empirical evaluation based on three image classification datasets validates our theoretical analysis, and demonstrates that HiFlash achieves satisfactory prediction performance for different levels of data heterogeneity and is communication-efficient compared with existing FL methods.

\bibliographystyle{unsrt}
\bibliography{reference}

% or


\ifCLASSOPTIONcaptionsoff
  \newpage
\fi

\end{document}